\documentclass[a4paper,11pt]{article}
\pdfoutput=1
\PassOptionsToPackage{hyperfootnotes=false}{hyperref}
\usepackage{float}
\usepackage{jheppub}
\usepackage[T1]{fontenc}
\usepackage{amsmath}
\usepackage{amssymb}
\usepackage{array}
\usepackage{fancyvrb}
\usepackage{booktabs}
\usepackage{etoolbox}
\usepackage{graphicx}
\usepackage{listings}
\let\paperteletypedefault\ttdefault
\usepackage[mono,varqu,noupquote]{inconsolata}
\let\ttdefault\paperteletypedefault
\usepackage{upquote}
\usepackage{microtype}
\usepackage{needspace}
\usepackage{placeins}
\usepackage{siunitx}
\usepackage{tabularx}
\usepackage[table]{xcolor}
\makeatletter
\let\paper@lstKeyInit\lst@Key@
\def\lst@Key@#1#2{  \ifstrequal{#1}{numbers}{}{\paper@lstKeyInit{#1}{#2}}}
\usepackage{lstlinebgrd}
\let\lst@Key@\paper@lstKeyInit
\patchcmd{\KV@lst@numbers}{none&}{none:}{}{}
\patchcmd{\KV@lst@numbers}{left&}{left:}{}{}
\patchcmd{\KV@lst@numbers}{right&}{right:}{}{}
\makeatother
\usepackage{tikz}
\usepackage[skins]{tcolorbox}
\usetikzlibrary{arrows.meta,positioning}
\graphicspath{{assets/}}
\let\papercontents\tableofcontents
\renewcommand{\tableofcontents}{  \begingroup
  \renewcommand{\baselinestretch}{1}\small
  \papercontents
  \endgroup}
\renewcommand{\afterTocRuleSpace}{\clearpage}
\newcolumntype{Y}{>{\raggedright\arraybackslash}X}
\DeclareRobustCommand{\software}[1]{\textrm{\upshape\textsc{#1}}}
\DeclareRobustCommand{\pyac}{\software{pyAmpliCol}}
\DeclareRobustCommand{\ac}{\software{AmpliCol}}
\DeclareRobustCommand{\mgfive}{\software{MadGraph5}}
\DeclareRobustCommand{\mgamc}{\software{MadGraph5\_aMC@NLO}}
\DeclareRobustCommand{\symbolica}{\software{Symbolica}}
\DeclareRobustCommand{\symjit}{\software{SymJIT}}
\DeclareRobustCommand{\spenso}{\software{Spenso}}
\DeclareRobustCommand{\idenso}{\software{Idenso}}
\DeclareRobustCommand{\rusticol}{\software{RustiCol}}
\DeclareRobustCommand{\ufoloader}{\software{ufo-model-loader}}
\DeclareRobustCommand{\comix}{\software{Comix}}
\DeclareRobustCommand{\pepper}{\software{Pepper}}
\DeclareRobustCommand{\recola}{\software{Recola}}
\DeclareRobustCommand{\pythia}{\software{Pythia}}
\DeclareRobustCommand{\sherpa}{\software{Sherpa}}
\DeclareRobustCommand{\herwig}{\software{Herwig}}
\DeclareRobustCommand{\madflow}{\software{MadFlow}}
\DeclareRobustCommand{\vincia}{\software{Vincia}}
\DeclareRobustCommand{\nnlojet}{\software{NNLOJET}}
\DeclareRobustCommand{\stripper}{\software{Stripper}}
\DeclareRobustCommand{\colorfulnnlo}{\software{CoLoRFulNNLO}}
\DeclareRobustCommand{\aloha}{\software{ALOHA}}

\DeclareRobustCommand{\feynrules}{\software{FeynRules}}
\DeclareRobustCommand{\openloops}{\software{OpenLoops}}
\DeclareRobustCommand{\madloop}{\software{MadLoop}}
\DeclareRobustCommand{\codex}{\software{Codex}}
\DeclareRobustCommand{\chatgpt}{\software{ChatGPT}}
\DeclareRobustCommand{\gnufortran}{\software{GNU Fortran}}
\DeclareRobustCommand{\maketool}{\software{make}}
\DeclareRobustCommand{\latexmk}{\software{latexmk}}
\DeclareRobustCommand{\pdflatex}{\software{pdfLaTeX}}
\definecolor{codetext}{HTML}{253247}
\definecolor{codekeyword}{HTML}{7026A3}
\definecolor{codecomment}{HTML}{627184}
\definecolor{codestring}{HTML}{A23C12}
\definecolor{codetype}{HTML}{005A7A}
\definecolor{codefunction}{HTML}{075ABB}
\definecolor{codebackground}{HTML}{F5F7FA}
\definecolor{codestripe}{HTML}{E7EDF5}
\definecolor{coderule}{HTML}{A6B6CA}
\lstdefinestyle{papercode}{
  basicstyle=\renewcommand{\ttdefault}{zi4}\ttfamily\fontsize{10}{11.5}\selectfont\color{codetext},
  keywordstyle=\color{codekeyword}\bfseries,
  keywordstyle=[2]\color{codefunction}\bfseries,
  commentstyle=\color{codecomment},
  stringstyle=\color{codestring},
  emphstyle=\color{codetype}\bfseries,
  columns=fullflexible,keepspaces=true,showstringspaces=false,
  upquote=true,tabsize=4,
  frame=none,backgroundcolor={},
  numbers=left,firstnumber=1,stepnumber=1,numberblanklines=true,
  numberstyle=\ttfamily\scriptsize\color{codecomment},numbersep=8pt,
  xleftmargin=20pt,xrightmargin=0pt,
  linebackgroundcolor={\ifodd\value{lstnumber}\color{codestripe}                       \else\color{codebackground}\fi},
  linebackgroundsep=3pt,
  linebackgroundwidth=\dimexpr\linewidth+3pt\relax,
  linebackgroundheight=8.5pt,linebackgrounddepth=3pt,
  aboveskip=0pt,belowskip=0pt,
  breaklines=true,breakatwhitespace=true,breakindent=1.5em
}
\lstdefinestyle{pythoncode}{
  style=papercode,language=Python,
  emph={Runtime,Path,TensorLibrary,Representation,TensorExpression,TensorName,
        TensorNetwork,LibraryTensor,ColorCorrelator,CorrelatorConfig,
        CorrelatedRequest,EmitGluon,Generator,ModelSource,ColorConfig,RunConfig,
        Decimal,Symbol,E,S,Rep,T,TN,AUTO},
  morekeywords=[2]{bis,mink,vector,dirac_gamma,to_network,execute,result_tensor,
                  components,evaluator,evaluate,evaluate_complex,evaluate_with_prec,
                  load,loads,read_text,evaluate_resolved,all_color,generate,
                  built_in_sm,available_color_correlations,evaluate_correlated_many}
}
\lstdefinestyle{shellcode}{
  style=papercode,language=bash,
  morekeywords={python3,python,pyamplicol,cd}
}
\tcbset{codepanel/.style={
  bicolor,
  colback=codebackground,colbacklower=white,
  colframe=coderule!60!white,
  colbacktitle=codebackground,coltitle=codetext,
  fonttitle=\small\sffamily,
  boxrule=0.4pt,titlerule=0pt,arc=1.5pt,
  boxsep=0pt,left=10pt,right=10pt,top=5pt,bottom=5pt,
  toptitle=5pt,bottomtitle=2pt,
  middle=5pt,
  segmentation style={solid,draw=coderule!45!white,line width=0.4pt},
  before skip=0.6\baselineskip,after skip=0.6\baselineskip
}}
\newtcolorbox{codecell}{codepanel}
\AtBeginDocument{
\newtcolorbox[use counter*=lstlisting]{notebookcell}[2]{
  codepanel,
  title={\textbf{Listing~\thetcbcounter}\enspace #1},
  label={#2}
}
}
\BeforeBeginEnvironment{table}{  \begingroup
  \setlength{\intextsep}{1.5\baselineskip}    \setlength{\abovecaptionskip}{0pt}  \setlength{\belowcaptionskip}{0.55\baselineskip}}
\AfterEndEnvironment{table}{\endgroup}
\makeatletter
\apptocmd{\float@endH}{  \ifdefstring{\@captype}{table}{\nointerlineskip}{}}{}{}
\makeatother

\title{\pyac{}: fast tree-level matrix elements}

\author[a]{Rikkert Frederix,}
\author[b,1]{Valentin Hirschi,\note{Corresponding author.}}
\author[b]{Lucien Huber,}
\author[f]{Shahriar Iravanian,}
\author[e]{Ben Ruijl,}
\author[c,d]{and Timea Vitos}
\affiliation[a]{Department of Physics, Lund University,\\
  S\"olvegatan 14A, SE-223 62 Lund, Sweden}
\affiliation[b]{Institute for Theoretical Physics, University of Bern,\\
  Sidlerstrasse 5, 3012 Bern, Switzerland}
\affiliation[c]{Department of Physics and Astronomy, Uppsala University,\\
  Box 516, 751 20 Uppsala, Sweden}
\affiliation[d]{Institute for Theoretical Physics, ELTE E\"otv\"os Lor\'and
  University,\\
  P\'azm\'any P\'eter s\'et\'any 1/A, H-1117 Budapest, Hungary}
\affiliation[e]{Ruijl Research, Chamerstrasse 117, 6300 Zug, Switzerland}
\affiliation[f]{Simrhythmia Research, Atlanta, GA, USA}

\emailAdd{rikkert.frederix@fysik.lu.se}
\emailAdd{valentin.hirschi@gmail.com}
\emailAdd{mail@lucien.ch}
\emailAdd{siravan@svtsim.com}
\emailAdd{ben@symbolica.io}
\emailAdd{timea.vitos@physics.uu.se}

\abstract{
Tree-level matrix elements sit at the heart of collider event generation, but their cost rises rapidly just where additional resolved radiation becomes most important. We present \href{https://mg5amcnlo.github.io/pyamplicol/}{\pyac}, a public generator aimed at extending the practical reach of colour-resolved matrix elements to higher multiplicities. It builds on the colour-ordered, off-shell-current strategy of \ac{}, providing fast access to individual leading-colour flows while retaining the option of next-to-leading or exact full colour. The latter introduces its own high-multiplicity bottleneck: the dense interference of factorially many colour orderings. \pyac{} addresses this contraction with a fast Fourier transform on the symmetric group, preserving the exact colour result while avoiding the direct pairwise sum. Representative full-colour benchmarks demonstrate substantial speedups over the reference calculations considered here. It supports Standard Model processes and a broad class of UFO models, selected or summed helicities, vectorised evaluation and controllable numerical precision. It further provides colour- and spin-correlated Born quantities and tree-level colour connections at arbitrary perturbative order, opening the same framework to local subtraction and other higher-order applications. Together, these capabilities connect high-multiplicity matrix elements to their use in event generation, merging, parton showers and precision calculations within a single generator.}


\hypersetup{
  pdftitle={\pyac{}: fast tree-level matrix elements},
  pdfauthor={Rikkert Frederix, Valentin Hirschi, Lucien Huber, Shahriar Iravanian, Ben Ruijl, Timea Vitos},
  pdfsubject={Tree-level matrix elements, colour decomposition and recursive amplitudes},
  pdfkeywords={QCD, matrix elements, colour decomposition, recursion, Monte Carlo}
}

\begin{document}
\maketitle
\raggedbottom

\section{Introduction}
\label{sec:introduction}

Tree-level scattering amplitudes are among the most frequently evaluated
objects in collider phenomenology.  They provide the hard processes for
leading-order event generation, the real-radiation terms of higher-order
calculations, and the fixed-order information used to improve parton showers.
The automation achieved by general-purpose generators has made a wide range of
Standard Model and beyond-the-Standard-Model reactions available to LHC
studies.

Alongside similar generators, \ac{}~\cite{Frederix:2026ejl} demonstrated that recursion relations and colour approximations, combined with optimised phase-space sampling, unlock high-multiplicity event generation. We present here \pyac{}, a re-implementation focused solely on the generation of optimised tree-level matrix elements, with three main objectives. First, use modern software development techniques to simplify the generation of and access to matrix elements. Second, generalise applicability to arbitrary processes and a vast range of generic models in the Universal \feynrules{} Output (UFO) format~\cite{Degrande:2011ua,Darme:2023jdn}, without performance regression in cases already covered by \ac{}. Third, extend the method to helicity-summed matrix elements with optimised exact colour treatment, delivering runtime performance surpassing that of existing exact matrix-element generators.

At high QCD multiplicity, amplitude construction and exact colour summation
present distinct computational challenges.  The number of Feynman diagrams
grows roughly factorially with the number of external particles.  For example, in the
diagram-based generator \mgamc{}~\cite{Alwall:2014hca}, first these contributions are enumerated and
only then shared wavefunctions that can be reused when evaluating the
helicity amplitudes are identified.  This reuse reduces the work, but does not eliminate
diagram enumeration.  Separately, in a decomposition into colour-ordered
partial amplitudes, a direct contraction of the dense colour-interference
matrix is quadratic in the factorially growing number of orderings.  This
second cost is independent of whether the partial amplitudes were built from
diagrams or recursion.  Berends--Giele-like improvements within \mgfive{}
likewise address amplitude construction separately from colour
summation~\cite{Lifson:2022Colour}.  The practical importance of this second
bottleneck is evident in the original \ac{} study: full-colour reweighting
grows faster than leading-colour event generation and eventually becomes the
limiting step~\cite{Frederix:2026ejl}.

Recursive constructions avoid the diagram-by-diagram expansion by reusing
off-shell currents~\cite{Berends:1987me,Caravaglios:1995cd,Kanaki:2000ey,
Moretti:2001zz,Duhr:2006iq}, and form the basis of modern
high-multiplicity generators such as \comix{}~\cite{Gleisberg:2008fv}, \pepper{}~\cite{Bothmann:2023gew}, and \ac{}~\cite{Frederix:2026ejl}.
For one fixed gluon ordering and helicity, binary Berends--Giele recursion has
cubic cost.  Explicit helicity sums can introduce exponential growth, while
enumerating all colour orderings remains factorial.  Efficient construction
of these ordered amplitudes therefore does not by itself remove the
factorial-squared cost of their direct colour
contraction~\cite{Duhr:2006iq,Frederix:2026ejl}.
Recursive off-shell currents also underpin the NLO matrix-element generator
\recola{}, where the construction is extended to the coefficients of
one-loop tensor integrals~\cite{Actis:2012qn,Actis:2016mpe}.

The factorial growth associated with colour orderings is not intrinsic to
exact colour.  At a fixed number of colours, orthogonal multiplet bases have
at most exponentially many basis vectors and reduce the final colour sum to
a diagonal contraction~\cite{Keppeler:2012ih}.  Direct off-shell recursion in
these bases has recently been developed for all-gluon tree amplitudes, with
exponential computational complexity~\cite{Bolinder:2025gbj}.  The improved
scaling comes at a cost, however: constructing the associated
multiplet-projected amplitudes is more expensive, making this approach less
efficient than methods based on fast Fourier transforms (FFTs) over the
phenomenologically relevant multiplicities for all-gluon processes studied in
ref.~\cite{ColourFFT:2026}.  Our approach therefore
retains colour-ordered amplitudes and complements fast recursion with FFTs
for exact colour summation.  Permutation symmetries reorganise the dense
interference sum into smaller matrix contractions without discarding
subleading-colour terms~\cite{ColourFFT:2026}.

The demand for high-multiplicity amplitudes is not confined to leading-order
predictions.  Multi-jet merging combines matrix elements with different
numbers of resolved partons with a parton shower, so that hard, well-separated
radiation is described by the appropriate fixed-order amplitude, the shower
describes the unresolved region, and the merging prescription controls double
counting.  This idea is realised in CKKW and CKKW-L
prescriptions~\cite{Catani:2001cc,Lonnblad:2001iq}, as well as MLM-based
approaches~\cite{Mangano:2006rw,Alwall:2007fs}.  Implementations include
tree-level merging in \pythia{}~8~\cite{Lonnblad:2011xx},
\sherpa{}~\cite{Hoeche:2009rj} and \herwig{}~\cite{Corcella:2000bw,Bellm:2015jjp}. Existing
NLO-matching and merging techniques, such as MEPS@NLO~\cite{Hoeche:2012yf} and FxFx~\cite{Frederix:2012ps}, also allow for complementing their higher-multiplicity contributions with fixed-order pure tree-level predictions.
Every additional tractable tree multiplicity extends the region in which the
hardest jets can be described directly by matrix elements.  Leading-order
matrix elements have been merged with parton showers for vector-boson
production with up to nine additional hard jets in parton-level
predictions~\cite{Hoche:2019flt,Brooks:2020mab}.  At the same time, the High-Luminosity LHC
(HL-LHC) programme is
expected to collect about $3\,\mathrm{ab}^{-1}$ at $14\,\mathrm{TeV}$,
substantially increasing the statistics in multi-jet final states and in the
high-energy tails where additional resolved radiation is
common~\cite{Azzi:2019yne,HSFPhysicsEventGeneratorWG:2020gxw}.  Fast tree-level
amplitudes at large multiplicity are therefore relevant both to the physics
reach and to the computing requirements of the next LHC data set.

Event generation and phase-space integration require the same matrix element
at many independent phase-space points.  This workload is naturally
parallelisable, and major event-generator projects increasingly use batched
evaluation on vector CPUs and accelerators.  This is visible in the
vector and GPU developments of \mgamc{}, in
\madflow{}, and in the portable \pepper{}
generator~\cite{Valassi:2025xfn,Frederix:2026dyl,Carrazza:2021gpx,Bothmann:2023gew}.  Vectorisation
complements amplitude-level improvements by applying the reduced calculation
to many points simultaneously. In
\pyac{}, native double-precision evaluation is vectorised
across phase-space points and supports single-instruction, multiple-data
(SIMD) execution.  This organisation provides a natural starting point for
extending the generated process output to GPUs.

Colour is particularly important in this high-multiplicity regime.  A
colour-flow decomposition writes a QCD amplitude as a sum of ordered partial
amplitudes multiplied by colour tensors~\cite{Paton:1969je,Mangano:1987xk,
DelDuca:1999rs,Maltoni:2002mq}.  The
leading-colour (LC) approximation retains the terms that dominate in the
large-$N_c$ limit, where $N_c$ is the number of colours, and avoids the complete
interference matrix between colour
orderings.  Its accuracy depends on the process and observable.  For the
processes studied in refs.~\cite{Frederix:2024uvy,Frederix:2026ejl}, the LC amplitudes
reproduce the dominant phase-space dependence sufficiently well to provide an
efficient proposal distribution for exact full-colour event generation.  The
\ac{} strategy first generates unweighted events using individual LC
contributions and subsequently restores full-colour accuracy by reweighting.
For the massless-QCD reactions studied through $2\to7$ partons, the acceptance
efficiency of this second unweighting step exceeds
$50\%$~\cite{Frederix:2024uvy}.  The
extension to multi-jet, $t\bar t+$jets, diboson+jets and Drell--Yan+jets
reactions retains high secondary efficiencies.  For the sampled subprocesses
with at most two quark lines, the measured event-generation cost grows
exponentially rather than factorially over the explored
range~\cite{Frederix:2026ejl}.  These observations motivate direct,
fast access to individual colour-ordered amplitudes, whether they are used as
an approximation, as integration channels, or as the starting point for an
exact-colour correction.

Closely related information appears in parton-shower applications.  A shower
describes the soft and collinear limits of successive emissions, whereas
matrix-element corrections modify those approximations so that resolved
configurations reproduce the relevant tree-level matrix
element~\cite{Seymour:1994df}.  Established examples include corrections to
vector-boson production in \pythia{}~\cite{Miu:1998ju} and
\herwig{}~\cite{Corcella:1999gs}.  In an LC shower, a
colour ordering is part of the state that determines which partons can radiate
coherently.  This connection is explicit in antenna showers and in
\vincia{}~\cite{Giele:2007di,Fischer:2016vfv}, where iterated
matrix-element corrections extend the correction beyond a single emission
to successively higher tree multiplicities~\cite{Giele:2011cb}.
Sector-shower merging exploits a unique shower history for each
colour-ordered final state, simplifying the reconstruction of histories at
high multiplicity~\cite{Brooks:2020mab}.
The colour dependence of the shower kernels can also be improved separately:
colour matrix-element corrections in the \herwig{}~7 dipole shower
retain interference between colour structures in the emission
probabilities~\cite{Platzer:2018pmd}.  They improve the finite-$N_c$ radiation
pattern rather than supplying the complete hard-emission matrix element.
Colour-resolved
matrix elements are thus useful not only because LC is often a good numerical
approximation, but because their structure matches the information carried by
the shower itself.

Fast high-multiplicity trees are also valuable in fixed-order calculations.
For a process with a tree-level Born contribution containing $n$ final-state
partons, the maximally real contribution at N$^k$LO involves tree-level matrix
elements with $n+k$ final-state partons.  This contribution is integrated over
the highest-dimensional phase space and, in local subtraction approaches,
evaluated together with counterterms.  It can consequently account for a
substantial part of the integration cost.  At NNLO, examples include antenna
subtraction as implemented in
\nnlojet{}~\cite{Gehrmann-DeRidder:2005btv,NNLOJET:2025rno},
sector-improved residue subtraction in
\stripper{}~\cite{Czakon:2010td,Czakon:2014oma}, nested soft-collinear
subtraction~\cite{Caola:2017dug}, \colorfulnnlo{}~\cite{DelDuca:2016ily},
and local analytic sector subtraction~\cite{Magnea:2018hab}.  The
soft and collinear limits of real radiation factorise onto reduced Born
amplitudes with colour and spin
insertions~\cite{Catani:1996vz,Catani:1999ss}.  Colour ordering and helicity
resolution therefore organise ingredients relevant to constructing the
colour- and spin-correlated Born quantities entering local subtraction
terms~\cite{Alioli:2013nda} (\emph{e.g.} as implemented in ref.~\cite{Buccioni:2019sur}).  Such insertions require amplitude interference
information beyond separate squared components.  \pyac{} retains these
interferences and supplies colour-connected and spin-projected Born quantities.
The colour construction is fully generic in perturbative order, not restricted
to NLO or NNLO: it recursively builds ordered emissions and quark-pair
splittings and exposes their compatible interferences.  An automatic option
registers a spanning set of tree-level soft-current colour structures through
any requested N$^k$LO order, which the user can project onto the combinations
needed by a subtraction scheme.

Subtraction calculations also expose demanding numerical regions.  The real
matrix element and its local approximation can be separately large while
their difference is small.  Stable kinematic parametrisations and analytic
forms control these cancellations~\cite{Bothmann:2024hgq}.  Higher-precision
evaluation complements them in validation, diagnostics and rescue calculations.
When symbolic expressions are retained,
\pyac{} permits the Python caller to choose the decimal precision
of the evaluation.  This includes intermediate as well as substantially higher precision,
without changing the requested process or its colour and helicity
decomposition.

Tree-level on-shell amplitudes also serve as building blocks for virtual
corrections, not only for real radiation. At one loop, the
Ossola--Papadopoulos--Pittau (OPP) method determines integral coefficients
by sampling the loop integrand at momenta that put selected propagators
on shell~\cite{Ossola:2006us}. Opening loops exposes tree-like structures,
allowing generators such as \madloop{}~\cite{Hirschi:2011pa} and
\openloops{}~\cite{Cascioli:2011va,Buccioni:2019sur} to reuse tree-level
techniques; \recola{} instead extends off-shell recursion directly to the
coefficients of one-loop tensor integrals~\cite{Actis:2012qn,Actis:2016mpe}.
Generalised unitarity makes the amplitude-level connection explicit:
loop amplitudes, also beyond one loop, can be reconstructed from suitable
cuts that factorise into products of on-shell tree amplitudes, summed over
the states crossing the cuts~\cite{Britto:2004nc,BernHuang:2011}.
Loop--tree duality (LTD) provides a related connection by expressing loop
integrals as phase-space integrals of cut, tree-like contributions, with
modified prescriptions for the remaining
propagators~\cite{Catani2008LoopTree,Capatti2019LoopTree}, with various
causal reorganisations~\cite{Capatti2020ManifestlyCausal,AguileraVerdugo2021Causal}
such as the cross-free family
representation~\cite{Capatti2023Threshold}.
Whereas generalised unitarity already works with amplitudes, these dual
representations are often constructed diagram by diagram. Making full use
of the shared currents in \pyac{} in this setting would require more
general amplitude-level relations. Progress in this direction includes a recursive construction
of one-loop amplitudes in massive scalar $\phi^3$ theory within causal
LTD~\cite{Kromin2022CausalAmplitudes}.
We hope that efficient tree-amplitude tools such as \pyac{} will encourage
further developments along these lines and help extend their practical
reach to higher multiplicities.

Despite the ubiquity of tree amplitudes, embedding a matrix-element provider
inside another phenomenological code often remains generator specific.  The
Les Houches event format standardises the exchange of generated
events~\cite{Alwall:2006yp}, the UFO format standardises model
information~\cite{Degrande:2011ua,Darme:2023jdn}, and the Binoth Les Houches Accord
standardises communication between Monte Carlo programs and one-loop
providers, including correlated Born requests~\cite{Alioli:2013nda}.  These
standards address adjacent stages of a calculation rather than providing one
common interface for repeated evaluation of colour-ordered tree amplitudes.  In
practice, applications may need to adapt to generated Fortran routines,
process-specific C++ libraries or Python wrappers with different conventions.
\pyac{} instead produces a single process output that can be evaluated directly from
Python, C, C++, Fortran and Rust.  This makes the same matrix element
available to event generators, subtraction frameworks and dedicated
phenomenological programs without changing its physical conventions.

We now turn to presenting the physics, implementation principles, validation
and performance of \pyac{}.  Installation instructions, examples and
the complete manual are maintained on the
\href{https://mg5amcnlo.github.io/pyamplicol/}{project website}~\cite{PyAmplicolDocs}.
In section~\ref{sec:ai-preface} we discuss the role of AI in developing
\pyac{} and writing this publication.
In section~\ref{sec:workflow} we present the implementation workflow,
including the recursion relations, contact interactions, colour sums and
correlated tree matrix elements. Section~\ref{sec:design} describes the
implementation design and maps its main software components.
In section~\ref{sec:results} we present numerical validation
and performance results, and we give our conclusions in
section~\ref{sec:conclusion}.  Appendix~\ref{app:benchmarks} contains further
benchmark comparisons.  Appendix~\ref{app:correlators} defines the correlated
tree matrix elements.  Appendix~\ref{sec:infrared-stability} examines their use
in local collinear and soft subtraction and the numerical stability of the
resulting differences.
Appendix~\ref{app:minimal-manual} gives a minimal installation and usage guide.

\section{Preface: the role of AI in this work}
\label{sec:ai-preface}
The \pyac{} project is an experiment in agentic scientific software development. One of the stated goals of the project was to confirm the utility of using agentic development in a real-world computational physics example.     

Starting from the physics ideas and experience embodied in \ac{}, we used \codex{}, primarily with GPT 5.6 Sol and earlier GPT 5.5 models, to port, generalise and test the implementation. We asked the model to work through several specialised subagents, while we set the physics requirements, inspected the results, designed validation tests and decided which changes were accepted. \codex{} also assisted in writing the first draft of this publication; however, the human authors have heavily edited and checked the resulting text. 

Without a doubt, large language models (LLMs) and, specifically, agentic development have disrupted computer programming and, more recently, mathematical research. Recent examples include finding a counterexample to the Jacobian conjecture and, more impressively, a (yet-to-be-confirmed) finite-time blow-up result for the Navier--Stokes equations. Nonetheless, we need to balance these successes with many attempts, failures, and poor-quality outputs. One of the key questions is what the characteristics of a successful AI-aided project are.

Successful AI-assisted work benefits from results that can be checked independently. For example, the counterexample to the Jacobian conjecture~\cite{gao2026jacobian} can be tested using a computer algebra system, or even by hand in this case. The reported Navier--Stokes finite-time blow-up result~\cite{openai2026navier}, accompanied by a Lean formalisation~\cite{openai2026lean}, illustrates this principle in mathematics.

In \pyac{}, independent matrix-element calculations and exact physics identities provide a different form of verification. The mature \ac{} implementation provided a strong reference for both physics conventions and numerical behaviour. We also generated full-colour tree-level matrix elements independently with \mgamc{}. We accepted representative changes after systematic numerical comparisons with an independent calculation, supplemented by internal checks of colour and helicity sums. These numerical tests do not constitute a formal proof of the implementation, but they similarly allow proposed results to be checked against criteria independent of the agent that produced them.

Moreover, expert supervision was necessary for the successful development of \pyac{}. Many rounds of instructions, correction and review were needed. The main development took place over roughly two weeks. As an intentional part of the experiment, all of the approximately half a million lines across the source code, tests and documentation were written directly by \codex{} rather than transcribed by us. We concentrated our effort on specifying the problem, providing physical insight, checking numerical results and challenging incorrect or unnecessarily complicated solutions. This division of labour is important: the speed of text and code production did not remove the need for expert supervision.

The experiment was resource intensive. At its peak, the work used access to nine \chatgpt{} Pro subscriptions, priced at CHF 200 per month each. A reconstruction from local session records, covering 3 July to 6 September 2026, reveals approximately 60.56 billion tokens across 430,598 requests. This covers the initial development, subsequent optimisation, numerical studies, documentation and work on this paper. 
Of 60.41 billion input tokens, 58.89 billion were cached (97.48\%); output accounted for a further 146.6 million tokens, including internal reasoning. Repeated records and histories inherited by subagents were counted only once. At the standard API prices of 6 September 2026, the recorded model mix and cache usage correspond to approximately USD 35,200~\cite{OpenAIAPIPricing}. 

\Needspace{3\baselineskip}
We paid for the subscriptions directly. The Swiss National Science Foundation (SNSF) declined to cover these costs, as they do not consider these to be compute costs. The scientific outcome of this experiment highlights the need for funding agencies and research institutions to support these tools as part of research computing.

\section{Overview of \pyac{}}
\label{sec:workflow}
\pyac{} generates tree-level matrix elements using off-shell recursion methods.
Therefore, the central object in its implementation is a colour-ordered off-shell current.  Model
information is first reduced to numerical rules for joining currents; a
requested scattering process is then assembled from those rules; and the
result is evaluated for batches of phase-space points.  This section follows
that sequence.  Details needed to run the program are kept in the online
documentation~\cite{PyAmplicolDocs}, and summarised in appendix~\ref{app:minimal-manual}.

\subsection{From a model to a matrix element}
\label{sec:model-workflow}

\begin{figure}[H]
  \centering
  \resizebox{\textwidth}{!}{\begin{tikzpicture}[
  node distance=5mm and 7mm,
  box/.style={draw=black!65, rounded corners=2pt, align=center,
              minimum height=9mm, inner xsep=6pt, fill=black!2,
              font=\small},
  component/.style={box, fill=blue!4},
  runtime/.style={box, fill=green!5},
  arrow/.style={-{Latex[length=2mm]}, thick, draw=black!70}
]
  \node[box] (ufo) {UFO\\model};
  \node[box, right=of ufo] (json) {JSON\\model};
  \node[component, right=of json] (process) {model processing\\and kernel derivation};
  \node[runtime, right=of process] (generation) {process\\generation};
  \node[box, right=of generation] (output) {process\\output};
  \node[runtime, right=of output] (evaluation) {batched\\evaluation};

  \draw[arrow] (ufo) -- (json);
  \draw[arrow] (json.east) -- (process.west);
  \draw[arrow] (process) -- (generation);
  \draw[arrow] (generation) -- (output);
  \draw[arrow] (output) -- (evaluation);
\end{tikzpicture}}
  \caption{Physics workflow in \pyac.  UFO input is first converted to the
  common JSON model description.  Model processing derives local current
  rules and kernels that can be reused across processes.  Process generation
  assembles them into the currents for the chosen scattering process; the
  resulting process output is evaluated repeatedly for batches of momenta.}
  \label{fig:workflow}
\end{figure}
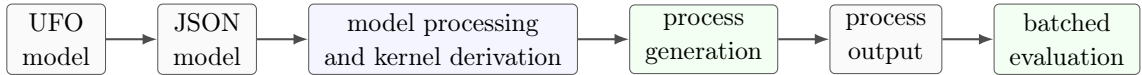

Figure~\ref{fig:workflow} summarises the workflow.  A model can be supplied as
a UFO directory~\cite{Degrande:2011ua,Darme:2023jdn}, as a
JSON description, or as one of the models included with \pyac.  A UFO is read
with \ufoloader{}~\cite{UFOLoader} and normalised
in memory to the same description used by the JSON route; no intermediate
JSON file is required.

During model processing, particle properties, parameters, couplings,
propagators and interaction tensors are checked and reduced to local current
rules.  The corresponding elementary calculations are called \emph{kernels}.
A processed model stores these rules and their symbolic expressions; a
\emph{prepared model} also stores compiled numerical routines for the kernels.
These process-independent ingredients can be saved and reused. Process generation
then combines these ingredients for chosen external states.

For a chosen process and perturbative order, the generator enumerates allowed
off-shell currents and shares identical ones between contributions.  It writes
self-contained process output containing the requested colour and helicity
coverage.  The same output can then be called from Python, C, C++, Fortran or
Rust.  These interfaces call \rusticol{}, the Rust component of \pyac{}
that efficiently loads and evaluates generated process output.  It is part of
\pyac{}, not a separate matrix-element generator, and accepts batches of
phase-space points for native double-precision evaluation.

Multiparticle labels allow a single process request to specify a whole set of
partonic subprocesses. Common model information and prepared kernels are
reused across the subprocesses, equivalent incoming or outgoing particle
orderings need only one evaluator, and the available CPU cores are shared
between concurrent generation tasks. Amusingly, even
\texttt{p p > 4*all} is a valid request: \texttt{p} denotes the chosen incoming
partons, while \texttt{all} contains every physical external particle of the
selected model. With the Standard Model, this can therefore request in one go
all allowed tree-level partonic subprocesses producing four SM particles in
proton--proton collisions, subject to the chosen flavour and coupling-order
settings, rather than just four-jet production.

\subsection{Off-shell-current recurrence}
\label{sec:recurrence}

\subsubsection{Organisation and reuse of the general recursion}
\label{sec:sharing-currents}

For a process with $N_{\rm ext}$ external particles, let $q_i$ be the
all-outgoing momentum of leg $i$: incoming momenta are taken with a minus
sign.  A colour sector specifies a colour tensor and the associated
ordering of coloured legs; each ordered list is called a \emph{colour word}.
Colour-singlet legs do not enter these words.  In each sector, one external
leg is reserved for the final contraction and called the \emph{closing leg}.
The remaining legs supply the one-particle currents: spinors, polarisation
vectors or scalar wavefunctions.

Composite currents are built in increasing numbers of external legs.
For a non-empty subset $I$ not containing the closing leg, define
$q_I=\sum_{i\in I}q_i$ and let $|I|$ denote its number of legs.
The current $J^a_{t,\Omega}(I)$ has outgoing particle species $t$ and
component $a$.  Its state label $\Omega$ records the external helicities in
$I$, spin or chirality, flavour and other conserved quantum numbers, colour
state and ordering, and coupling order.

\Needspace{8\baselineskip}
Except for the final amputated current described below, composite currents
with $|I|\geq2$ obey
\begin{align}
 J^a_{t,\Omega}(I)={}&P_t{}^a{}_b(q_I)
 \sideset{}{'}\sum_{\substack{I=A\mathbin{\dot\cup}B\\A,B\ne\varnothing}}
 \sum_{\substack{v:t_A t_B\to t\\\Omega_A,\Omega_B}}
 \sigma_v(A,B)\,w_v(A,B)\,g_v
 \nonumber\\
 &\qquad\times V_v{}^b{}_{cd}(q_A,q_B)
 J^c_{t_A,\Omega_A}(A)J^d_{t_B,\Omega_B}(B).
 \label{eq:generic-current}
\end{align}
Here $P_t$ is the propagator of species $t$\footnote{For a non-propagating
auxiliary current, $P_t$ is the identity.  The rewriting of supported contact
interactions into binary joins is discussed in
section~\ref{sec:contact-decomposition}.}.  The label $v$ denotes an allowed
three-point vertex, and $V_v$ is its colour-stripped rule joining input
species $t_A,t_B$ into $t$.
The factors $g_v$, $w_v$ and $\sigma_v$ are respectively its coupling,
local colour-decomposition weight and fermion-reordering sign.
The indices $a,b,c,d$ denote Lorentz, Dirac or scalar components, with
repeated component indices summed.  The symbol $A\mathbin{\dot\cup}B$
denotes a disjoint partition; the primed sum counts each allowed partition
and vertex orientation once, with identical-field combinatorial factors
included in the local rule.  Only joins and input states $\Omega_A,\Omega_B$
compatible with $\Omega$, the model and the chosen colour ordering are
retained.  This is the Berends--Giele construction~\cite{Berends:1987me}
with the particle content and local rules of the supplied model.

Contributions to the same current are summed before applying its propagator
once.  At the final stage, $|I|=N_{\rm ext}-1$, this propagator is omitted:
the resulting amputated current is contracted with the closing-leg
wavefunction to give the partial amplitude.  The closing-leg choice is an
internal convention, not a physical restriction.

Currents with the same complete labels $(I,t,\Omega)$ are evaluated once
and reused wherever they occur, including across colour sectors and
helicity assignments of the other legs.  Identical colour-stripped joins
can likewise be shared even when they enter with different local weights.
This structural reuse is exact; further relations, such as a sign change
under an exchange of inputs, may follow from the vertex algebra.
An optional search for equal, opposite or zero currents uses high-precision
numerical probes and independent verification points.  These additional
relations are numerically tested rather than algebraically proven; the
search is off by default and enabled with
\texttt{--numerical-current-reuse}.
\pyac{} retains the shared-current construction of
\ac{}~\cite{Frederix:2026ejl}, extending its model rules and colour
projections and removing its restriction to at most three open quark lines.

\subsubsection{The example \texorpdfstring{$d\bar d\to Zgg$}{d dbar to Zgg}}
\label{sec:zgg-example}
Consider, with $p_i$ the momentum of external particle $i$,
\begin{equation}
  d(p_1)\,\bar d(p_2)\longrightarrow Z(p_3)\,g(p_4)\,g(p_5).
  \label{eq:zgg-process}
\end{equation}
We use all-outgoing momenta
\begin{equation}
  q_1=-p_1,\qquad q_2=-p_2,\qquad q_i=p_i\quad(i=3,4,5),
  \qquad \sum_{i=1}^{5}q_i=0 .
  \label{eq:all-outgoing}
\end{equation}
At tree level an exact colour decomposition of the amplitude ${\cal M}_h$ is
\begin{equation}
  {\cal M}_h=g_s^2\!\!\sum_{\sigma\in S_2}
  \left(T^{a_{\sigma(4)}}T^{a_{\sigma(5)}}\right)_{i_2}^{\ \bar\imath_1}
  A_h\bigl(2,\sigma(4),\sigma(5),1;3_Z\bigr).
  \label{eq:zgg-colour-decomposition}
\end{equation}
Here $g_s$ is the strong coupling, with $\alpha_s=g_s^2/(4\pi)$,
and $T^a$ are fundamental $SU(3)$ generators.  The indices $a_4,a_5$ label
the gluon colours, while $i_2,\bar\imath_1$ label the outgoing quark and
antiquark obtained by crossing the incoming legs.  The label $h$ denotes
the complete assignment of external helicities.  We
use $\operatorname{tr}(T^aT^b)=T_R\delta^{ab}$ with $T_R=1/2$.
$S_2$ permutes the gluon labels 4 and 5.
The electroweak $Z d\bar d$ coupling is included in the partial amplitude
$A_h$.  Its arguments give the colour word along the quark line, followed
after the semicolon by the colour-singlet leg $3_Z$.  The two words are
\begin{equation}
  c_0=(2,4,5,1),\qquad c_1=(2,5,4,1).
  \label{eq:zgg-words}
\end{equation}

For $c_0$, choose leg 1 as the closing leg.  Let $G(I)$ and $Q(I)$ denote
gluon and down-quark currents with component and state labels suppressed.
The quark currents are row spinors grown from the outgoing quark
$Q(2)=\bar u_{h_2}(q_2)$ towards the outgoing antiquark on leg 1, whose
column spinor $v_{h_1}(q_1)$ closes the amplitude; $h_i$ is the helicity of
leg $i$.  The other one-leg currents are the polarisation vectors $G(4)$,
$G(5)$ and $Z(3)$.  Composite currents contain subsets of $\{2,3,4,5\}$
whose coloured legs are consecutive in $c_0$: gluon 4 precedes gluon 5,
whereas the $Z$ can be emitted anywhere along the quark line.

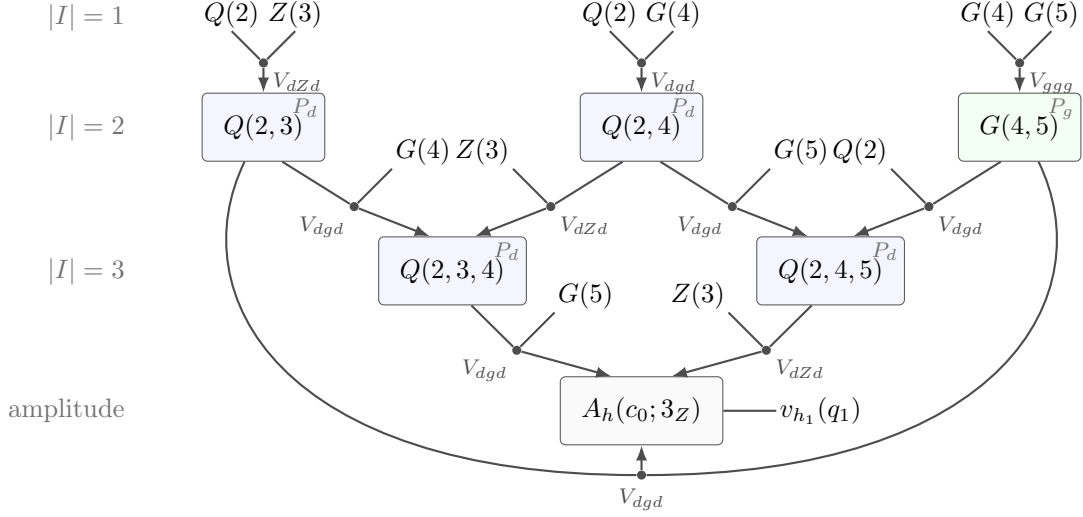
\begin{figure}[htbp]
  \centering
  \begin{tikzpicture}[
  box/.style={draw=black!65, rounded corners=2pt, align=center,
              minimum height=9mm, inner xsep=6pt, fill=black!2,
              font=\small},
  current/.style={box, inner xsep=8pt},
  quark/.style={current, fill=blue!4},
  gluon/.style={current, fill=green!5},
  amplitude/.style={box, fill=black!2, inner xsep=8pt},
  tag/.style={font=\scriptsize, text=black!55, inner sep=1pt},
  vertex/.style={circle, fill=black!70, inner sep=0pt, minimum size=3.4pt},
  rule/.style={font=\scriptsize, text=black!70, inner sep=1pt},
  src/.style={font=\small, inner sep=1pt},
  header/.style={font=\small, text=black!55, anchor=east},
  link/.style={thick, draw=black!70},
  arrow/.style={-{Latex[length=2mm]}, thick, draw=black!70}
]
    \node[quark]     (q23)  at (3,4.4)     {$Q(2,3)$};
  \node[quark]     (q24)  at (8,4.4)     {$Q(2,4)$};
  \node[gluon]     (g45)  at (13,4.4)    {$G(4,5)$};
  \node[quark]     (q234) at (5.5,2.5)   {$Q(2,3,4)$};
  \node[quark]     (q245) at (10.5,2.5)  {$Q(2,4,5)$};
  \node[amplitude] (amp)  at (8,0.65)    {$A_h(c_0;3_Z)$};
  \foreach \n/\p in {q23/d,q24/d,g45/g,q234/d,q245/d}
    \node[tag, anchor=north east] at ([xshift=-1.5pt,yshift=-1pt]\n.north east) {$P_\p$};

    \node[vertex] (j2)  at (3,5.25)     {};
  \node[vertex] (j3)  at (8,5.25)     {};
  \node[vertex] (j1)  at (13,5.25)    {};
  \node[vertex] (j4)  at (4.2,3.35)   {};
  \node[vertex] (j5)  at (6.8,3.35)   {};
  \node[vertex] (j6)  at (9.2,3.35)   {};
  \node[vertex] (j7)  at (11.8,3.35)  {};
  \node[vertex] (j8)  at (6.35,1.45)   {};
  \node[vertex] (j9)  at (9.65,1.45)   {};
  \node[vertex] (j10) at (8,-0.2)   {};

    \draw[link] (j2) -- ++(-0.42,0.42) node[src, anchor=south] {$Q(2)$};
  \draw[link] (j2) -- ++( 0.42,0.42) node[src, anchor=south] {$Z(3)$};
  \draw[link] (j3) -- ++(-0.42,0.42) node[src, anchor=south] {$Q(2)$};
  \draw[link] (j3) -- ++( 0.42,0.42) node[src, anchor=south] {$G(4)$};
  \draw[link] (j1) -- ++(-0.42,0.42) node[src, anchor=south] {$G(4)$};
  \draw[link] (j1) -- ++( 0.42,0.42) node[src, anchor=south] {$G(5)$};
  \draw[link] (j4) -- ++( 0.5,0.5) node[src, anchor=south west] {$G(4)$};
  \draw[link] (j5) -- ++(-0.5,0.5) node[src, anchor=south east] {$Z(3)$};
  \draw[link] (j6) -- ++( 0.5,0.5) node[src, anchor=south west] {$G(5)$};
  \draw[link] (j7) -- ++(-0.5,0.5) node[src, anchor=south east] {$Q(2)$};
  \draw[link] (j8) -- ++( 0.5,0.5) node[src, anchor=south west] {$G(5)$};
  \draw[link] (j9) -- ++(-0.5,0.5) node[src, anchor=south east] {$Z(3)$};
  \draw[link] (amp.east) -- ++(0.7,0) node[src, anchor=west] {$v_{h_1}(q_1)$};

    \draw[link] ([xshift= 2.5mm]q23.south)  -- (j4);
  \draw[link] ([xshift=-2.5mm]q24.south)  -- (j5);
  \draw[link] ([xshift= 2.5mm]q24.south)  -- (j6);
  \draw[link] ([xshift=-2.5mm]g45.south)  -- (j7);
  \draw[link] ([xshift= 2.5mm]q234.south) -- (j8);
  \draw[link] ([xshift=-2.5mm]q245.south) -- (j9);
  \draw[link] ([xshift=-2.5mm]q23.south) .. controls (1.9,2.2)  and (3.2,-0.2)  .. (j10);
  \draw[link] ([xshift= 2.5mm]g45.south) .. controls (14.1,2.2) and (12.8,-0.2) .. (j10);

    \draw[arrow] (j2)  -- (q23);
  \draw[arrow] (j3)  -- (q24);
  \draw[arrow] (j1)  -- (g45);
  \draw[arrow] (j4)  -- ([xshift=-3mm]q234.north);
  \draw[arrow] (j5)  -- ([xshift= 3mm]q234.north);
  \draw[arrow] (j6)  -- ([xshift=-3mm]q245.north);
  \draw[arrow] (j7)  -- ([xshift= 3mm]q245.north);
  \draw[arrow] (j8)  -- ([xshift=-4mm]amp.north);
  \draw[arrow] (j9)  -- ([xshift= 4mm]amp.north);
  \draw[arrow] (j10) -- (amp.south);

    \node[rule, below right=1pt and 1pt of j2]  {$V_{dZd}$};
  \node[rule, below right=1pt and 1pt of j3]  {$V_{dgd}$};
  \node[rule, below right=1pt and 1pt of j1]  {$V_{ggg}$};
  \node[rule, below left =1pt and 1pt of j4]  {$V_{dgd}$};
  \node[rule, below right=1pt and 1pt of j5]  {$V_{dZd}$};
  \node[rule, below left =1pt and 1pt of j6]  {$V_{dgd}$};
  \node[rule, below right=1pt and 1pt of j7]  {$V_{dgd}$};
  \node[rule, below left =1pt and 1pt of j8]  {$V_{dgd}$};
  \node[rule, below right=1pt and 1pt of j9]  {$V_{dZd}$};
  \node[rule, below=2pt of j10]               {$V_{dgd}$};

    \node[header] at (1.3,5.85) {$|I|=1$};
  \node[header] at (1.3,4.4)  {$|I|=2$};
  \node[header] at (1.3,2.5)  {$|I|=3$};
  \node[header] at (1.3,0.65) {amplitude};
\end{tikzpicture}
  \caption{Shared-current construction of the partial amplitude
  $A_h(c_0;3_Z)$ for $c_0=(2,4,5,1)$, closing on antiquark leg 1.
  Dots are binary joins labelled by their colour-stripped rules; short
  external lines supply the one-leg currents.  Blue and green boxes collect
  quark and gluon currents, respectively, with propagators marked in their
  corners.  Rows follow the number $|I|$ of collected legs.  The final box
  sums the three closing joins without a propagator and contracts with
  $v_{h_1}(q_1)$.  The currents $Q(2,3)$, $Q(2,4)$ and $G(4,5)$ each feed
  two later joins.  The ten joins are those of
  eqs.~\eqref{eq:zgg-currents}--\eqref{eq:zgg-closure}.}
  \label{fig:zgg-recursion}
\end{figure}

Write $V_{dgd}$, $V_{ggg}$ and $V_{dZd}$ for the quark--gluon, three-gluon
and $Zd\bar d$ rules, with their two input currents in square brackets.
The QCD rules have their strong couplings stripped, as in
eq.~\eqref{eq:zgg-colour-decomposition}, while $V_{dZd}$ retains its
electroweak coupling.  The symbols $P_d$ and $P_g$ denote the down-quark
and gluon propagators; their momenta, fermion-reordering signs and local
colour-order weights are left implicit.  Figure~\ref{fig:zgg-recursion}
shows the resulting construction.

The two-leg currents are
\begin{align}
 G(4,5) & =P_g V_{ggg}\bigl[G(4),G(5)\bigr],
 \nonumber\\
 Q(2,3) & =P_d V_{dZd}\bigl[Q(2),Z(3)\bigr],
 \nonumber\\
 Q(2,4) & =P_d V_{dgd}\bigl[Q(2),G(4)\bigr].
 \label{eq:zgg-currents}
\end{align}
$Q(2,5)$ is absent because gluon 4 must be met before gluon 5 along the
quark line of $c_0$, and no vertex couples the $Z$ to gluons.  Each
three-leg current receives two contributions, added before the single
propagator is applied,
\begin{align}
 Q(2,3,4) & =P_d\left\{
 V_{dgd}\bigl[Q(2,3),G(4)\bigr]
 +V_{dZd}\bigl[Q(2,4),Z(3)\bigr]\right\},
 \nonumber\\
 Q(2,4,5) & =P_d\left\{
 V_{dgd}\bigl[Q(2,4),G(5)\bigr]
 +V_{dgd}\bigl[Q(2),G(4,5)\bigr]\right\},
 \label{eq:zgg-three-leg-currents}
\end{align}
while $\{2,3,5\}$ is excluded like $Q(2,5)$ and $\{3,4,5\}$ has no vertex.
At the last stage the four-leg quark current is kept without its propagator
and contracted with the closing-leg spinor.  Its admissible partitions
$\{2,3,4,5\}=A\mathbin{\dot\cup}B$ have $B=\{5\}$, $\{3\}$ or $\{4,5\}$,
since $B=\{4\}$ would need the excluded $Q(2,3,5)$ and a $B$ containing the
$Z$ together with a gluon has no vertex.  With all particles outgoing, the
partial amplitude for $c_0$ is therefore
\begin{align}
  A_h(c_0;3_Z)={}&\Bigl\{
  V_{dgd}\bigl[Q(2,3,4),G(5)\bigr]
  +V_{dZd}\bigl[Q(2,4,5),Z(3)\bigr]
  \nonumber\\
  &\quad
  +V_{dgd}\bigl[Q(2,3),G(4,5)\bigr]\Bigr\}\,v_{h_1}(q_1).
  \label{eq:zgg-closure}
\end{align}
Substituting eqs.~\eqref{eq:zgg-currents} and
\eqref{eq:zgg-three-leg-currents} into eq.~\eqref{eq:zgg-closure} and using
the linearity of the vertex rules in their arguments unfolds the recursion
completely,
\begin{align}
  A_h(c_0;3_Z)={}&\biggl\{
  V_{dgd}\Bigl[P_d V_{dgd}\bigl[P_d V_{dZd}[Q(2),Z(3)],G(4)\bigr],G(5)\Bigr]
  \nonumber\\
  &\;+V_{dgd}\Bigl[P_d V_{dZd}\bigl[P_d V_{dgd}[Q(2),G(4)],Z(3)\bigr],G(5)\Bigr]
  \nonumber\\
  &\;+V_{dZd}\Bigl[P_d V_{dgd}\bigl[P_d V_{dgd}[Q(2),G(4)],G(5)\bigr],Z(3)\Bigr]
  \nonumber\\
  &\;+V_{dZd}\Bigl[P_d V_{dgd}\bigl[Q(2),P_g V_{ggg}[G(4),G(5)]\bigr],Z(3)\Bigr]
  \nonumber\\
  &\;+V_{dgd}\Bigl[P_d V_{dZd}\bigl[Q(2),Z(3)\bigr],
  P_g V_{ggg}\bigl[G(4),G(5)\bigr]\Bigr]
  \biggr\}\,v_{h_1}(q_1).
  \label{eq:zgg-diagrams}
\end{align}
The five terms are the five colour-ordered Feynman diagrams of $c_0$: with
both gluons attached directly to the quark line, the $Z$ is emitted before
gluon 4, between the two gluons or after gluon 5; with the gluons first
fused by the three-gluon vertex, the $Z$ is emitted after or before the
resulting off-shell gluon.  Evaluated diagram by diagram, they would need
fifteen vertex evaluations per helicity structure; the recursion of
eqs.~\eqref{eq:zgg-currents}--\eqref{eq:zgg-closure} needs ten, namely
$1+1+1+2+2+3$: three evaluations are saved because $G(4,5)$, $Q(2,3)$ and
$Q(2,4)$ are each computed once and used twice, and two more because the two
contributions to each three-leg current in
eq.~\eqref{eq:zgg-three-leg-currents} are summed before the final joins with
$G(5)$ and $Z(3)$, which are then evaluated once instead of twice.

Attaching the external helicities reproduces the ``one word'' column of
Table~\ref{tab:zgg-current-census}.  With two helicities for each massless
quark and gluon and three for the massive $Z$, the sources are the nine
wavefunctions on legs 2--5 and the two spinors $v_{h_1}(q_1)$, 11 in total.
A composite current exists once per helicity assignment of its legs:
$G(4,5)$, $Q(2,3)$ and $Q(2,4)$ have $4$, $6$ and $4$ copies, the 14 currents
with $|I|=2$; $Q(2,3,4)$ and $Q(2,4,5)$ have $12$ and $8$, the 20 with
$|I|=3$; and the amputated four-leg current in the braces of
eq.~\eqref{eq:zgg-closure} has $24$, for $11+14+20+24=69$ stored currents
receiving $14+40+72=126$ binary contributions, one per join and per copy.
The process has $48$ physical helicity assignments, one for each amputated
current and each state of $v_{h_1}(q_1)$.  Twenty-four vanish for massless
down quarks because the helicities on the fermion line must be chirality
compatible: only one state of $v_{h_1}(q_1)$ closes a given amputated
current, leaving 24 structurally non-zero closures.

The second word $c_1=(2,5,4,1)$ is built in the same way with gluons 4 and 5
interchanged.
Joint construction shares all one-particle wavefunctions, the six copies of
$Q(2,3)$, which does not see the gluons, and the four copies of the
two-gluon current, because the ordered three-gluon rule is antisymmetric in
its two arguments, so that the stripped currents satisfy $G(5,4)=-G(4,5)$; a
single copy is stored and the sign is carried by the local colour-order
weight of each word.  Hence the ``both words'' column: $14+14-10=18$
currents with $|I|=2$, $40$ and $48$ with $|I|=3,4$, where nothing is
shared, $117$ stored currents and $2\times126-10=242$ binary contributions.

Of these, 32 quark--gluon-to-quark joins reuse the same colour-stripped
evaluation across the two words: the eight helicity copies of
$V_{dgd}[Q(2),G(4,5)]$ and the 24 copies of
$V_{dgd}[Q(2,3),G(4,5)]$.  Their word-dependent weights are applied
separately, leaving $242-32=210$ distinct evaluations.
Compared with constructing the words separately, joint construction thus
reduces the stored currents from $2\times69=138$ to 117 and the local
evaluations from $2\times126=252$ to 210, savings of about 15\% and 17\%,
respectively.

\begin{table}[htbp]
  \centering
  \caption{Current counts for $d\bar d\to Zgg$ with the full helicity basis in
  the built-in Standard Model of \pyac{}.  ``One word'' means $c_0$ or
  $c_1$; ``both words'' means their joint construction.  Sources are external
  wavefunctions and closures are final closing-leg contractions.  Evaluations count
  distinct colour-stripped three-point joins.}
  \label{tab:zgg-current-census}
  \small
  \renewcommand{\arraystretch}{1.12}
  \begin{tabular}{l r r}
    \toprule
    Quantity & One word & Both words \\
    \midrule
    One-particle wavefunctions & 11 & 11 \\
    Currents with $|I|=2,3,4$ & $14,\ 20,\ 24$ & $18,\ 40,\ 48$ \\
    Stored currents, including sources & 69 & 117 \\
    Binary current contributions & 126 & 242 \\
    Distinct colour-stripped evaluations & 126 & 210 \\
    Structurally non-zero closures & 24 & 48 \\
    \bottomrule
  \end{tabular}
\end{table}

\subsection{Decomposition of contact interactions}
\label{sec:contact-decomposition}

\subsubsection{The four-gluon interaction}

The four-gluon vertex provides a useful example of turning a contact
interaction into cubic steps~\cite{Draggiotis:2002hm,Duhr:2006iq}.  With
$A^a_\mu$ the gluon field and $f^{abc}$ the $SU(3)$ structure constants, we write the Yang--Mills Lagrangian ${\cal L}_{\mathrm{YM}}$ in terms of the field strength $F^a_{\mu\nu}$ and of the bilinear $X^a_{\mu\nu}$ that carries its non-Abelian part,
\begin{equation}
  {\cal L}_{\mathrm{YM}}=-\frac14F^a_{\mu\nu}F^{a\mu\nu},
  \qquad
  F^a_{\mu\nu}=\partial_\mu A^a_\nu-\partial_\nu A^a_\mu
  +g_sX^a_{\mu\nu},
  \qquad X^a_{\mu\nu}=f^{abc}A^b_\mu A^c_\nu .
\end{equation}
The quartic term is then
\begin{equation}
  {\cal L}_4=-\frac{g_s^2}{4}X^a_{\mu\nu}X^{a\mu\nu}.
  \label{eq:four-gluon-term}
\end{equation}
Introduce an antisymmetric adjoint auxiliary field $B^a_{\mu\nu}$ with the Lagrangian
\begin{equation}
  {\cal L}_{\mathrm{aux}}=\frac14B^a_{\mu\nu}B^{a\mu\nu}
   -\frac{g_s}{2}B^a_{\mu\nu}X^{a\mu\nu}.
  \label{eq:four-gluon-auxiliary}
\end{equation}
Its algebraic equation of motion is $B^a_{\mu\nu}=g_sX^a_{\mu\nu}$;
substitution into eq.~\eqref{eq:four-gluon-auxiliary} reproduces
eq.~\eqref{eq:four-gluon-term}.  Since $B$ has no derivative term, it carries
no physical pole and introduces no propagating state.  More explicitly,
using $g_{\mu\nu}=\mathrm{diag}(1,-1,-1,-1)$ and Feynman rules from
$\exp(iS)$, the cubic $BAA$ vertex $\Gamma$ and algebraic $BB$ contraction
$D_B$ are
\begin{align}
  \Gamma^{eab}_{\alpha\beta;\mu\nu}
    &=-i g_s f^{eab}
      (g_{\alpha\mu}g_{\beta\nu}-g_{\alpha\nu}g_{\beta\mu}),
      \nonumber\\
  D_B^{ef;\alpha\beta,\gamma\delta}
    &=i\delta^{ef}
      (g^{\alpha\gamma}g^{\beta\delta}-g^{\alpha\delta}g^{\beta\gamma}).
  \label{eq:four-gluon-cubic-rules}
\end{align}
Here $\Gamma$ couples $B^{e\alpha\beta}$ to $A^{a\mu}$ and $A^{b\nu}$.
We use only the independent components of $B$: each auxiliary pair is
restricted to $\alpha<\beta$, and likewise $\gamma<\delta$, including in
internal sums.  Connecting two $\Gamma$ vertices with $D_B$ and summing the
three distinct pairings of four gluons gives the original four-gluon rule.
Each pairing therefore requires two binary joins and an intermediate current
with six independent Lorentz components.  The ordinary three-gluon vertex
is unchanged.

This example uses an auxiliary field to expose a useful tensor
factorisation.  The implementation need not introduce such a field in the
Lagrangian: it can derive the same intermediate components directly from
the original vertex.  We next describe this more general construction and
then return to the four-gluon interaction to make the connection explicit.

\subsubsection{General factorisation of contact interactions}
\label{sec:general-contact-decomposition}

Consider one colour-stripped term of an $n$-point rule,
$V_{\alpha_1\ldots\alpha_n}(p_1,\ldots,p_n)$, where $\alpha_i$ collects the
Lorentz or spin components of leg $i$.  Choose leg $r$ as the result leg,
leaving $\alpha_r$ open and contracting the input currents $J_i$ on all other
legs.  The resulting vertex contribution, before attaching its propagator,
is
\begin{align}
  [K_r]_{\alpha_r}={}&g_V
  \sum_{\{\alpha_i:i\ne r\}}
  V_{\alpha_1\ldots\alpha_n}(p_1,\ldots,p_n)
  \prod_{i\ne r}(J_i)_{\alpha_i},
  \nonumber\\[-2pt]
  &p_r=-\sum_{i\ne r}p_i .
  \label{eq:npoint-current}
\end{align}
All momenta are incoming to the local vertex.  The component sums are in a
fixed basis, with metric and spinor contractions included in $V$; $g_V$
contains the coupling and any scalar prefactor, including the local colour
weight when working in a chosen colour ordering.  The result leg is not
necessarily an initial or final particle in the scattering process: it
specifies which index remains open.  Each inequivalent result field and
required input ordering must be accounted for.

To evaluate the contact in stages, divide its input legs into non-empty,
disjoint sets $S$ and $T$, with
$S\mathbin{\dot\cup}T=\{1,\ldots,n\}\setminus\{r\}$.  Write
$\alpha_S=(\alpha_i)_{i\in S}$ and $\alpha_T=(\alpha_j)_{j\in T}$ for the
grouped component indices.  At fixed momenta, $\alpha_S$ and
$(\alpha_T,\alpha_r)$ can be regarded as the row and column indices of a
finite matrix, which always admits an exact factorisation; the issue is
whether the auxiliary space is compact.  In practice, UFO rules already
specify sums of products and contractions of familiar tensors, such as
metrics, momenta and Dirac matrices.  We exploit the explicit tensor
structure supplied by the UFO and exact relations among its components;
we do not perform a general search for compact tensor decompositions.
Where no compact construction is available, supported generic colour-singlet
contacts are instead evaluated by retaining the input currents and momenta
until the final join, without claiming a reduction in contraction cost.
An exact factorisation takes the form
\begin{equation}
  V_{\alpha_S,\alpha_T,\alpha_r}
  =\sum_A L_{A;\alpha_S}\,R_{\alpha_r;A,\alpha_T},
  \label{eq:contact-factorisation}
\end{equation}
where $A$ labels auxiliary components, $L$ contracts the currents on $S$,
and $R$ attaches those on $T$ to produce the result.  Momentum arguments are
suppressed here and below, not assumed absent.  The first step forms the
non-propagating auxiliary current
\begin{equation}
  X_A(S)=\sum_{\alpha_S}L_{A;\alpha_S}\prod_{i\in S}(J_i)_{\alpha_i},
  \label{eq:contact-auxiliary-current}
\end{equation}
and the second closes it against the remaining inputs,
\begin{equation}
  [K_r]_{\alpha_r}
  =g_V\sum_A\sum_{\alpha_T}R_{\alpha_r;A,\alpha_T}\,X_A(S)
   \prod_{j\in T}(J_j)_{\alpha_j},
  \label{eq:contact-reconstruction}
\end{equation}
which reproduces eq.~\eqref{eq:npoint-current} by direct substitution.

An explicit, though not necessarily compact, construction uses the open
indices themselves as a complete component basis.  Let $A$ run over the
joint values of $(\alpha_T,\alpha_r)$ and set
\begin{equation*}
  L_{A;\alpha_S}=V_{\alpha_S,A},
  \qquad
  R_{\alpha_r;A,\alpha_T}=\delta_{A,(\alpha_T,\alpha_r)},
\end{equation*}
where the Kronecker delta is an identity between component labels, not a
Lorentz metric.  The auxiliary current then stores the partial contraction
of $V$ with the inputs on $S$.  This establishes exactness, but not efficiency:
the complete basis has $\prod_{j\in T\cup\{r\}}d_j$ components if index
$\alpha_j$ takes $d_j$ values.  For derivative interactions, $L$ and hence
$X_A(S)$ may also depend on momenta in $T$ and on $p_r$.  Such a current cannot
in general be reused using only the data on $S$.

Repeated splits organise the $n-1$ inputs into a binary tree with $n-2$
two-input joins and $n-3$ intermediate currents.  A balanced tree has at most
$\lceil\log_2(n-1)\rceil$ levels.  These are counts of joins, not scalar
operations: compact intermediate spaces must still be found if the
factorisation is to reduce the contraction cost.  None of the auxiliary
states introduces a propagator or a new pole.

For supported momentum-independent four-point rules, \pyac{} implements
the component-basis construction directly.  It chooses two input legs from
the colour structure: two structure constants sharing an adjoint index fix
the pair, while for a colour-singlet rule it takes the first two non-result
legs.  \spenso{}~\cite{SpensoSoftware} translates the UFO Lorentz expression
into a tensor network and contracts it with symbolic components of the two
input currents.  The remaining input and result indices stay open, giving
16 candidate components for four vectors.  \pyac{} uses
\symbolica{}~\cite{SymbolicaSoftware} to remove exact zeros and keep a single
representative of components which are equal or opposite.  It records the
map needed to reconstruct the full tensor; the second join contracts that
tensor with the remaining input current.  Thus the components are derived
from the UFO rule rather than a model-specific hand-written decomposition.
This reduction exploits exact component relations, not a general search for
a minimum-rank factorisation.

For example, let $g_6$ be the complete rule factor of a six-scalar contact
term and choose scalar 6 as the result.  Auxiliary subscripts now list the
combined legs, since scalars carry no open index.  One balanced tree is
\begin{equation}
  X_{12}=J_1J_2,\qquad X_{45}=J_4J_5,\qquad
  X_{345}=J_3X_{45},\qquad K_{\phi_6}=g_6X_{12}X_{345}.
  \label{eq:six-scalar-tree}
\end{equation}
It contains four joins and three non-propagating scalar currents.  The
coupling and perturbative order enter only in the last join.  When this
labelled tree is entered in the recursion, the last-join coefficient is
divided by the exact number of equivalent input assignments generated by
the tree, preventing identical fields from overcounting the source vertex.
The corresponding ten-scalar construction has eight joins and seven
intermediate currents.

Returning to the four-gluon interaction, choose leg 4 as the result.  In the
conventions of eq.~\eqref{eq:four-gluon-cubic-rules}, its full rule is
\begin{equation}
  (\Gamma_4)^{a_1a_2a_3a_4}_{\mu_1\mu_2\mu_3\sigma}
  =-i g_s^2\sum_{p=(i,j;k)\in\mathcal P}\mathcal C_p
    (g_{\mu_i\mu_k}g_{\mu_j\sigma}
     -g_{\mu_i\sigma}g_{\mu_j\mu_k}),
  \qquad
  \mathcal C_p=f^{e a_i a_j}f^{e a_k a_4},
  \label{eq:four-gluon-factorisation}
\end{equation}
where $\mathcal P=\{(1,2;3),(1,3;2),(2,3;1)\}$ lists the three pairings,
and the repeated adjoint index $e$ is summed.  Expand each colour factor in
the chosen colour tensors $c_\omega$ as
$\mathcal C_p=\sum_\omega w_p^{(\omega)}c_\omega$; the coefficients
$w_p^{(\omega)}$ are the corresponding colour-decomposition weights.  For
each pairing the first join forms an antisymmetric tensor, and the second
gives its contribution to the current in colour channel $\omega$:
\begin{align}
  X^{(ij)}_{\rho\sigma}
    &=J_{i\rho}J_{j\sigma}-J_{i\sigma}J_{j\rho},
    \nonumber\\
  K^{(\omega)}_{4,\sigma}
    &=-i g_s^2\sum_{p=(i,j;k)\in\mathcal P}
       w_p^{(\omega)}X^{(ij)}_{\rho\sigma}J_k^\rho .
  \label{eq:antisymmetric-contact}
\end{align}
Explicitly, collect the covariant components of an input gluon current in
the column $\mathbf J_i=(J_{i0},J_{i1},J_{i2},J_{i3})^{\mathsf T}$.
These are supplied by the preceding recursion, or by a polarisation vector
for a one-particle input.  The intermediate tensor is the matrix
\[
  \mathbf X^{(ij)}
  =\mathbf J_i\mathbf J_j^{\mathsf T}-\mathbf J_j\mathbf J_i^{\mathsf T}
  =\begin{pmatrix}
    0       & x_{01}  & x_{02}  & x_{03} \\
    -x_{01} & 0       & x_{12}  & x_{13} \\
    -x_{02} & -x_{12} & 0       & x_{23} \\
    -x_{03} & -x_{13} & -x_{23} & 0
  \end{pmatrix},
  \qquad
  x_{\rho\sigma}=J_{i\rho}J_{j\sigma}-J_{i\sigma}J_{j\rho}.
\]
Here ${}^{\mathsf T}$ denotes transpose without complex conjugation.
The final contraction in eq.~\eqref{eq:antisymmetric-contact} uses
$J_k^\rho=g^{\rho\lambda}J_{k\lambda}$, so the metric enters when this
remaining current is attached.

This is the four-gluon construction used by \pyac{}, with signs and overall
normalisations distributed between the UFO coupling, Lorentz rule and local
colour weights according to their conventions.  Of the 16 candidate
components of each $X^{(ij)}$, four vanish and six are the negatives of the
other six.  The automatic reduction therefore recovers the same six stored Lorentz
components and two binary joins per pairing as the auxiliary-field
construction.  Their computational structure is equivalent; no larger
intermediate tensor has to be stored simply because the decomposition was
derived generically.

The contracted form in eq.~\eqref{eq:contact-auxiliary-current} is used for
scalar contact trees and the checked four-point and Higgs effective field theory (HEFT) classes.  Other
supported indexed colour-singlet contacts, including momentum-dependent
four-point rules and higher-point interactions, use the input-retaining
construction described above, organised as a balanced binary tree.
\spenso{} derives the final-join component formulae from the original UFO
expression, with the result-field indices left open as in
eq.~\eqref{eq:npoint-current}.

For example, consider six distinguishable colour-singlet vector fields with
the Lorentz tensor
\[
  V_{\mu_1\ldots\mu_6}
    =g_{\mu_1\mu_2}g_{\mu_3\mu_4}g_{\mu_5\mu_6}.
\]
Choosing leg 6 as the result, \spenso{} contracts the five input currents to give the result current on leg 6 with open index $\mu_6$,
$K_{6,\mu_6}=g_V(J_1\mathbin{\cdot}J_2)
(J_3\mathbin{\cdot}J_4)J_{5,\mu_6}$ automatically, where the dot denotes a
Lorentz contraction.  The current generic
implementation derives this final expression; it does not automatically
replace the intermediate states by two scalar products and a vector.

The common requirement is exact algebraic reconstruction of the source
rule, rather than a prescription tied to a particular model.  For compact
decompositions, \pyac{} uses construction-specific symbolic proofs: it
checks the allowed tensor structure and retains the colour-index connections,
component reconstruction, permutation signs and identical-field factors
needed for every inequivalent result field.  These checks do not rely on
agreement at sampled numerical points.  The generic packed construction,
by contrast, evaluates the original rule directly.  This framework allows
further decompositions to be added with their reconstruction checks without
changing the binary recursion; it does not automatically prove arbitrary
proposed factorisations.  HEFT is one example within its present scope:
the $Hggg$ and $Hgggg$ contacts of a CP-even scalar $H$ admit checked auxiliary
constructions, including their momentum dependence.  The currently supported
UFO structures are summarised in table~\ref{tab:ufo-coverage}.
\spenso{} supplies the tensor contractions and component formulae;
section~\ref{sec:spenso} gives a concrete code example.

\subsection{Supported UFO input}

Table~\ref{tab:ufo-coverage} summarises the supported UFO input; the online
documentation lists every condition and diagnostic.  The
tests cover the Standard Model, HEFT for a
CP-even scalar with $Hgg$, $Hggg$ and $Hgggg$ interactions~\cite{Dawson:1990zj},
a massless spin-two field coupled to scalars, and
colour-singlet scalar contacts with three through ten fields.

\begin{table}[htbp]
  \centering
  \caption{Supported UFO input in \pyac{}~1.0.0 with \texttt{ufo-model-loader}~1.0.0.
  Features in the last column are rejected at model preflight or compilation
  with an explicit diagnostic; conditional cases are summarised here and
  spelled out, with every condition and diagnostic code, in the documentation
  page \url{https://mg5amcnlo.github.io/pyamplicol/user/ufo-model-coverage/},
  which is kept in step with the code.}
  \label{tab:ufo-coverage}
  \footnotesize
  \renewcommand{\arraystretch}{1.08}
  \begin{tabularx}{\textwidth}{>{\raggedright\arraybackslash\bfseries}p{0.15\textwidth}YY}
    \toprule
    Feature & Supported & Rejected or restricted \\
    \midrule
    Input
      & UFO directories (trusted Python), serialised JSON models, compiled
        model files and prepared kernel bundles; restriction cards for
        UFO and JSON sources.
      & \texttt{ufo-model-loader} older than 1.0.0; restriction cards on
        built-in or precompiled models. \\
    Fields
      & Scalars, Dirac fermions, vectors and massless spin two (validated);
        massive spin two (experimental).
      & Majorana and fermion-number-violating fermions, spin $3/2$ and
        higher; ghosts and Goldstone bosons only internally. \\
    Colour
      & One $SU(3)$ with representations $1$, $3$, $\bar3$ and $8$;
        external coloured legs are triplet fermions or octet vectors.
      & Sextets, further colour groups, $N_c\neq3$; the totally symmetric colour tensor $d^{abc}$,
        the antisymmetric $\epsilon$ colour tensor and sextet tensors. \\
    Lorentz
      & Identity, $\gamma^\mu$, $\gamma^5$, chiral projectors,
        $\sigma^{\mu\nu}$, metric and leg momenta; slashed momenta in
        propagators; a fixed registry of scalar functions.
      & Levi-Civita and charge-conjugation tensors, form factors, functions
        outside the registry. \\
    Three-point vertices
      & Any spin combination with colour structure $1$, Kronecker $\delta$, $T^a$ or
        $f^{abc}$, momentum dependent or not.
      & $d^{abc}$; one- and two-point terms are ignored. \\
    Four-point vertices
      & Colour-singlet contacts with any Lorentz structure; the
        two-$f^{abc}$ gauge contact; the scalar-HEFT $Hggg$ family.
      & Other coloured contacts, e.g.\ four-fermion operators. \\
    Higher-point vertices
      & Colour-singlet contacts of any valence (tested to ten scalars);
        the scalar-HEFT $Hgggg$ family.
      & Every other coloured contact. \\
    Propagators
      & Standard scalar, Dirac, vector and spin-two kernels with fixed
        widths: Feynman gauge for massless and unitary gauge for massive
        vectors, Goldstones absorbed; model-supplied propagators that
        lower exactly.
      & Complex-mass scheme; other gauge choices. \\
    Parameters
      & External and internal, real or complex; runtime cards for external
        parameters with derived couplings recomputed.
      & Custom function bodies; $\epsilon$-expanded values keep their
        finite part; counterterm vertices are dropped. \\
    Processes
      & $2\to N$ tree amplitudes, with $N$ the number of final-state particles; leading colour (LC), next-to-leading colour (NLC) and full colour; FFT contraction
        in the trace or adjoint basis.
      & Loop-induced channels (omitted with a warning). \\
    \bottomrule
  \end{tabularx}
\end{table}

\subsection{Efficient colour sums}
\label{sec:colour-sums}

For a chosen colour basis and fixed external helicity assignment $h$, write a
tree amplitude and its colour-summed square as
\begin{equation}
  {\cal M}_h=\sum_\alpha C_\alpha A_{h,\alpha},
  \qquad
  {\cal B}_h\equiv\sum_{\text{colours}}|{\cal M}_h|^2=A_h^\dagger G A_h,
  \qquad
  G_{\alpha\beta}=\sum_{\text{colours}}C_\alpha^*C_\beta .
  \label{eq:colour-metric}
\end{equation}
The indices $\alpha,\beta$ run over the elements of the colour basis: $C_\alpha$ is a trace, adjoint half-ladder or open-quark-line colour tensor, $A_{h,\alpha}$ its
ordered partial amplitude, $A_h$ is the vector of these amplitudes, and $G$
is the exact colour-overlap matrix.  The basis
is adapted to the external coloured particles~\cite{Paton:1969je,
Mangano:1987xk,DelDuca:1999rs,Maltoni:2002mq}.

Let $N_{\rm ord}$ be the number of retained orderings.  A dense evaluation of
eq.~\eqref{eq:colour-metric} takes $\mathcal O(N_{\rm ord}^2)$ work, even if
all partial amplitudes have already been obtained by fast recursion.  For a
pure-gluon amplitude with $N_g$ external gluons, including the incoming ones,
the unreduced trace set
has $N_{\rm ord}=(N_g-1)!$ after fixing one leg by cyclicity; the adjoint
decomposition fixes two end legs and has $N_{\rm ord}=(N_g-2)!$.
Direct colour squaring therefore costs $\mathcal O([(N_g-1)!]^2)$ or
$\mathcal O([(N_g-2)!]^2)$, respectively.  These are counts of direct
interference operations~\cite{Lifson:2022Colour,ColourFFT:2026}.
\pyac{} adapts its FFT basis to the process. For Yang--Mills gluon trees
and single-insertion HEFT amplitudes, it uses the two-anchor
Del Duca--Dixon--Maltoni (DDM) decomposition when the model's colour tensors
satisfy the required tree identities; colourless spectators do not enter
$N_g$. Quark amplitudes retain fundamental chains or their products, while
other interactions keep the original colour representation unless a DDM
reduction is established. This choice preserves the exact colour sum but
does not reduce the basis for every process.

At leading colour (LC), a
physical flow---one allowed connection pattern among the external colour
lines---or a sum of flows can be requested directly.  At next-to-leading
colour (NLC) and full colour, the ordered partial amplitudes are contracted
with the colour-overlap matrix at the requested accuracy.  The precise NLC prescription is
process dependent:  for amplitudes with only gluons as coloured legs, in the trace
or DDM basis, NLC retains terms at the basis-wide
leading power $P$ and first subleading power $P-2$ of $N_c$ in the overlap matrix $G$; lower powers are dropped, \textit{i.e.}, a strict $1/N_c$ expansion is enforced. For processes with an open quark line, the colour factors in the overlap matrix $G$ that do not contain a leading $P$ or subleading power $P-2$ in $N_c$ are set to zero, while the full coefficients, including all subleading powers without truncation, are kept for the other terms~\cite{Frederix:2024uvy}.

Two exact transforms reduce repeated work in this contraction.  In some
open-quark-line bases, parts of $G$ are invariant under $k$ independent
exchanges, which form the Abelian group $(\mathbb Z_2)^k$.  Ordering each orbit
by its group elements permits a
  decomposition into the group characters through the Walsh--Hadamard
transform~\cite{Walsh:1923,Shanks:1969}.
The current release uses verified blocks with $k=2$ or 3.  This is a change
of basis, not a colour approximation.  It is applied automatically by the
direct recurrence contraction when such a block is present.

For a complete orbit of $m$ permuted gluon labels under the symmetric group
$S_m$, with $m=N_g-1$ for the complete trace set or $m=N_g-2$ for the DDM set, overlaps
depending only on the relative permutation $p^{-1}q$, with $p,q\in S_m$,
define a group convolution acting on the amplitude vector.  A non-Abelian
fast Fourier transform (FFT) reorganises it into matrices associated with
irreducible representations of $S_m$, labelled by Young
diagrams~\cite{DiaconisRockmore1990,ClausenBaum1993,Maslen1998,ColourFFT:2026}.
For one complete orbit there are $m!$ amplitude values.  The Fourier matrices
collectively contain the same number of entries, but only matrices with the
same Young diagram couple.  The colour matrices in this representation are
independent of momenta and helicities and are prepared once.  At each
evaluation, a fast amplitude transform and the smaller matrix contractions
replace the direct sum over $(m!)^2$ pairs~\cite{ColourFFT:2026}.
\pyac{} first verifies the complete group action, currently for $m\leq10$.
All pairs of verified orbit channels, including their interference, use the
transform, while incomplete pieces remain in the
untransformed contraction, so the requested NLC or full-colour result is
unchanged. For NLC and full-colour recurrence or on-the-fly output,
FFT with the adaptive basis is enabled by default where supported;
otherwise generation falls back to direct contraction. The options
\texttt{--fft adjoint} and \texttt{--fft trace} explicitly request the
adaptive or original basis, respectively, while
\texttt{--color-contraction direct} disables FFT. LC, compiled, eager and
correlated outputs retain direct contraction. Since the direct recurrence
already uses the Walsh reduction where applicable, the additional
symmetric-group transform can have a small effect in some multi-quark-line
processes even when it is important for complete all-gluon permutation
orbits.

\subsection{Correlated tree matrix elements}
\label{sec:correlated-design}
The partial amplitudes of section~\ref{sec:colour-sums} also provide the
correlated Born quantities needed in local subtraction. These require colour
and spin insertions in a reduced Born amplitude, rather than its ordinary
squared matrix element. \pyac{} retains the complex partial amplitudes and
replaces the colour-overlap matrix $G$ of eq.~\eqref{eq:colour-metric} by the
matrix of a requested colour insertion.  The user
specifies these insertions at generation and selects them by name at
evaluation.  A shorthand constructs all ordered emission and splitting
histories through a chosen order and registers their compatible bra--ket
interferences.  This includes radiation from emitted gluons and quark pairs,
not just dipole products; the resulting nonminimal set spans tree-level
soft-current colour structures.  Their conventions, including the distinction between
connection order and colour accuracy, are set out in
appendix~\ref{app:correlators}.

For spin correlations, the user declares which vector particles may be
contracted and supplies their four-vectors at runtime.  A supplied vector
replaces the corresponding external polarisation literally; several vectors
act jointly on the same amplitude.  This gives both spin projections and
mixed spin--colour contractions.  Multiple requests can share one call,
including a batch of phase-space points: colour insertions with the same
spin assignment reuse the same amplitudes, while different assignments reuse
evaluation stages whose inputs are unchanged.

Correlated output supports LC, NLC and full colour.  It keeps the complete
colour basis and applies the selected approximation to the inserted colour
matrix, following the existing colour-accuracy conventions.  It uses direct
contraction and generic compiled amplitudes, with binary64 evaluation in
\rusticol{} through the C, C++, Fortran and Rust interfaces.  Python additionally
provides arbitrary-precision and explicitly selected double-double evaluation;
spin vectors supplied as decimal numbers retain their precision on input.
This correlated calculation does not use FFT, on-the-fly
or recurrence execution, physical-helicity zero and parity reductions, or
numerically inferred current identities.  These restrictions retain the
interferences and arbitrary vector dependence required by the insertions;
they do not change ordinary uncorrelated generation or evaluation.  The
\href{https://mg5amcnlo.github.io/pyamplicol/correlators.html}{online manual}
gives the complete interface and runnable examples.

\section{Implementation details}
\label{sec:design}

\label{sec:structure}

The public entry points of \pyac{} are a command-line program and a Python interface for
selecting a model, steering process generation and loading the resulting
output.  The resulting matrix elements can be called from Python, C,
C++, Fortran or Rust.  These five interfaces are thin wrappers around \rusticol{}, the Rust backend
that efficiently loads and evaluates the generated processes.
The public interfaces return squared matrix elements, resolved by helicity
and, at LC, by colour flow.  At NLC and full colour the colour contraction is
already performed.  These interfaces also supply correlated
Born insertions as described in section~\ref{sec:correlated-design}, retaining
their complex interference without exposing the full array of partial amplitudes.

When constructing the \rusticol{} backend, the Python interface allows for
four arrangements to evaluate the current construction of
section~\ref{sec:recurrence}. Where a mode supports the requested workload,
these arrangements calculate the same amplitude.
\begin{enumerate}
\item \emph{recurrence}: This mode executes a
    compact sequence of shared-current joins directly. Compared to the other modes, 
    this generally keeps setup and memory moderate.
\item \emph{compiled}: This mode combines larger portions of
    the process calculation into process-specific stage evaluators. This mode
    favours repeated evaluation.
\item \emph{eager}: This mode stores a flat table of prepared model-kernel
    calls, their inputs and their combination rules, without compiling
    process-wide stages.
\item \emph{on-the-fly}: This mode prepares only the selected
    helicity and colour contributions on first use. This mode is particularly
    useful when only a small part of the available helicity or colour coverage will be needed.
\end{enumerate}

The default \emph{recurrence} mode delivers good performance in general, and the typical user should not have to worry about switching away from it. The \emph{compiled} mode is relevant mostly to stress-test improvements in \symjit{} and the \emph{on-the-fly} mode lets users defer the choice of which outputs to pay the
generation cost for until they are needed. 
More importantly, at generation time, the user must choose between two \texttt{lc\_flow\_layout} options:
\begin{itemize}
    \item \texttt{lc\_flow\_layout = "topology-replay"} (default): optimises the evaluation of a selected colour flow, summed over all physical helicity configurations.
    \item \texttt{lc\_flow\_layout = "all-flow-union"}: optimises the evaluation of all individual colour flows for a selected helicity configuration.
\end{itemize}
All workloads are supported by either mode, but they do not show identical performance. More information can be found in the
\href{https://mg5amcnlo.github.io/pyamplicol/user/runtime-and-selectors/}{live runtime documentation} of \pyac{}.
 
Internally, \pyac{} relies on three main software libraries to
efficiently generate and/or evaluate matrix elements: \symbolica{}, \symjit{}, and
\spenso{}. The following subsections describe how these libraries are used.

\begin{figure}[htbp]
  \centering
  \resizebox{\textwidth}{!}{\begin{tikzpicture}[
  box/.style={draw=black!55, line width=0.45pt, rounded corners=2pt,
    align=center, text width=3.75cm, minimum height=1.8cm,
    inner xsep=5pt, inner ysep=6pt,
    font=\fontsize{9}{11}\selectfont, fill=black!2},
  python/.style={box, fill=blue!4},
  native/.style={box, fill=green!5},
  data/.style={-{Latex[length=2mm]}, line width=0.65pt, draw=black!65},
  call/.style={data, dashed, draw=blue!45!black},
  edge label/.style={font=\footnotesize, text=black!75, fill=white,
    inner sep=2pt, align=center}
]
  \node[box] (input) at (1.95,0)
    {\textbf{Model input}\\UFO / JSON / built-in};
  \node[python] (model) at (7.05,0)
    {\textbf{Model processing}\\normalise local rules\\prepare kernels};
  \node[box] (compiled) at (12.15,0)
    {\textbf{Compiled model}\\rules and expressions\\
     optional prepared kernels};

  \node[python, inner ysep=3pt] (cli) at (1.95,-3.4)
    {\textbf{Command Line}\\\textbf{Interface}\\commands and run cards\\\texttt{pyamplicol}};
  \node[python] (generator) at (7.05,-3.4)
    {\textbf{Python Generator}\\process planning\\generation control};
  \node[native] (builder) at (12.15,-3.4)
    {\textbf{Rust generation}\\current construction\\execution plans};

  \node[box] (output) at (7.05,-5.85)
    {\textbf{Process output}\\process data and plans\\kernels and API examples};

  \node[python] (python) at (1.95,-8.3)
    {\textbf{Python Runtime}\\load and profile\\batched evaluations};
  \node[native] (runtime) at (7.05,-8.3)
    {\textbf{\rusticol{} evaluator}\\shared native evaluation\\colour / helicity sums};
  \node[python] (api) at (12.15,-8.3)
    {\textbf{Native-code API}\\\mbox{C, C++, Fortran, Rust}\\language wrappers};

    \draw[data] (input) -- (model);
  \draw[data] (model) -- (compiled);
  \draw[data] (compiled.south west) --
    node[edge label, right=2pt] {reuse} (generator.north east);
  \draw[call] (cli.north east) --
    node[edge label, left=2pt] {prepare} (model.south west);
  \draw[call] (cli) -- (generator);
  \draw[call, {Latex[length=2mm]}-{Latex[length=2mm]}]
    (generator) -- (builder);
  \draw[data] (generator) -- node[edge label, right=2pt] {write} (output);

    \draw[data] (output) -- node[edge label, right=2pt] {load} (runtime);
  \draw[call] (cli) -- node[edge label, right=3pt] {evaluate\\or profile} (python);
  \draw[call] (python) -- (runtime);
  \draw[call] (api) -- (runtime);
  \draw[data] (output.east) --
    node[edge label, above=2pt] {\texttt{API/} examples}
    (12.15,-5.85) -- (api.north);

  \draw[data] (2.7,-9.8) -- (3.6,-9.8);
  \node[anchor=west, font=\footnotesize, text=black!75] at (3.75,-9.8)
    {model or process data};
  \draw[call] (8.4,-9.8) -- (9.3,-9.8);
  \node[anchor=west, font=\footnotesize, text=black!75] at (9.45,-9.8)
    {calls and returns};
\end{tikzpicture}}
  \caption{High-level component map of \pyac{}.  Blue boxes denote Python
  components or public interfaces, green boxes native Rust computation, and
  grey boxes model or process data.  Solid arrows show data exchange; dashed
  arrows show calls and returns.  A compiled model contains reusable model
  rules, not a complete scattering-process evaluator; a prepared model also
  contains compiled numerical kernels.  The process output retains the plans
  and kernels needed for evaluation, so the original model bundle and
  generator are no longer required.  Its optional \texttt{API/} examples
  call the Python or native interfaces; the wrappers and \rusticol{} library are supplied
  with the installed package.  The standard double-precision path is shown.}
  \label{fig:code-structure}
\end{figure}
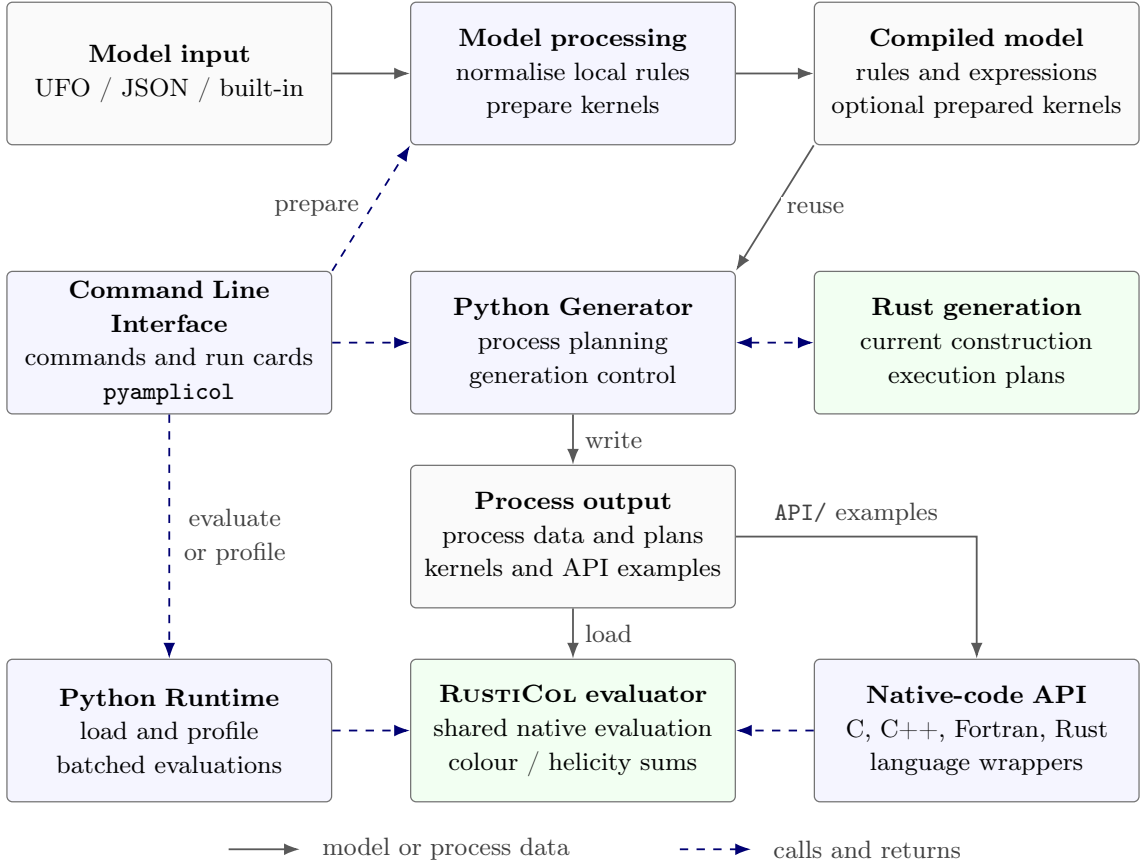

Figure~\ref{fig:code-structure} locates the main software components and the
data exchanged between them.  Model processing uses \symbolica{} and \spenso{} for
symbolic expressions and \symjit{} to prepare numerical kernels.
The Python \texttt{Generator} coordinates
process construction, using model rules and native Rust builders.  The Rust
generation backend and the \rusticol{} evaluator are two roles of the same
\rusticol{} implementation, not independent libraries.

Detailed implementation descriptions, together with complete
installation, generation and API references, dependency versions and supported
platforms, are maintained in the
\href{https://mg5amcnlo.github.io/pyamplicol/}{online documentation}~\cite{PyAmplicolDocs}.

\subsection{Symbolic manipulations with \symbolica{}}
\label{sec:symbolica}
\symbolica{}~\cite{SymbolicaSoftware} is a framework for symbolic and numerical
computations, implemented in Rust with Python bindings. It combines efficient
multivariate polynomial arithmetic, a condensed memory model for expressions, and fast pattern matching with an intuitive
interface for manipulating large symbolic expressions. A key feature for
\pyac{} is its ability to rapidly construct numerical evaluators from symbolic
expressions.

\symbolica{} can rewrite multiple nested expressions into a single optimised evaluator.
It first rewrites the expression symbolically, using Horner schemes to reduce the number of multiplications,
and it detects and removes common subtrees, so that common parts are only evaluated once.
Then it turns the tree representation into a sequence of simple native instructions, such as addition, multiplication,
and function calls. Next, the sequence is optimised by iteratively removing common pairs~\cite{Kuipers:2013pba}.
The optimisation procedure is designed to be fast for large expressions, and avoids excessive memory usage.

The resulting evaluator can be called with many numerical types, including finite fields, floating-point
numbers of different sizes (double precision at 53 bits, double-double precision at 106 bits, and arbitrary precision),
and complex numbers.

\pyac{} uses these strengths during model processing and process
generation: exact couplings and component expressions obtained from model
tensors are simplified and then translated into numerical kernels.  This is
especially useful for UFO input, where a vertex cannot be assumed to belong to
a short list of hard-coded forms.

\Needspace{17\baselineskip}
An evaluator can be compiled to machine code using various backends. One option is
C++ with inline assembly. The inline assembly substantially reduces compilation time,
as expensive rewriting operations of \texttt{G++}/\texttt{Clang} can be skipped, without loss of performance.

Another backend is \symjit{} (see section~\ref{sec:symjit}), which is a just-in-time compiler for symbolic expressions.
This is the preferred backend and offers a significant speedup over the C++ backend.

In the following example, \symbolica{} creates an optimised evaluator for a nested expression.

\begin{notebookcell}{Building a numerical evaluator from a symbolic expression.}{lst:symbolica-symjit}
\begin{lstlisting}[style=pythoncode]
from symbolica import *
x, y, z, f, g = S('x', 'y', 'z', 'f', 'g')

expr = x + Symbol.PI + x.cos() + f(g(x+1), x*2)
f_body = y**2 + z**2*y**2
g_body = y + 5

ev = expr.evaluator([x], functions=[
    FunctionDefinition(f, [y, z], f_body),
    FunctionDefinition(g, [y], g_body),
])
\end{lstlisting}
\end{notebookcell}

It generates a linear list of instructions, reusing registers, that can be translated into machine code:

\begin{Verbatim}[fontsize=\small]
('fun', ('temp', 0), cos, [], [('param', 0)], False)
('add', ('temp', 1), [('param', 0), ('const', 0)], 0)
('add', ('temp', 1), [('const', 1), ('temp', 1)], 0)
('mul', ('out', 0), [('param', 0), ('const', 2)], 0)
('mul', ('temp', 1), [('temp', 1), ('temp', 1)], 0)
('mul', ('out', 0), [('out', 0), ('out', 0)], 0)
('mul', ('out', 0), [('temp', 1), ('out', 0)], 0)
('add', ('temp', 1), [('temp', 1), ('out', 0)], 0)
('add', ('out', 0), [('param', 0), ('const', 3), ('temp', 0), ('temp', 1)], 0)
\end{Verbatim}

Below we show three ways to run the evaluator: (1) just-in-time compilation with SIMD support for double-precision numbers, (2) evaluation using double-double precision (106-bit precision), and (3) evaluation using arbitrary precision.

\begin{notebookcell}{Calling the evaluator.}{lst:symbolica-evaluation}
\begin{lstlisting}[style=pythoncode]
ev.evaluate([1., 2., 3., 4.]) # JIT backend using SIMD

ev.evaluate_with_prec([Float(1, decimal_digits=32)], 32) # double-double

ev.evaluate_with_prec([Float(1, decimal_digits=100)], 100)
\end{lstlisting}
\end{notebookcell}

An interactive version of the example above can be found on the \href{https://symbolica.io/gallery}{Symbolica gallery}.

When process output retains symbolic expressions, the user has access to the precision-controlled \texttt{evaluate\_with\_prec} path.
The caller selects a decimal
precision appropriate to the calculation, including intermediate precision
and substantially higher precision for diagnostic checks.  For ordinary
matrix elements this path uses guarded multiprecision, also around 32 digits;
correlated calls additionally offer explicit double-double arithmetic with
31-digit transfers, as illustrated in appendix~\ref{app:minimal-manual}.
On-the-fly output and symmetric-group FFT colour contraction are
currently double-precision only.  For the supported higher-precision path, the
process definition, helicity and colour decomposition do not change.
The calculation uses the coefficients and current relations retained during
generation.  The optional numerical reuse described in
section~\ref{sec:sharing-currents} is off by default, so a process generated
without it provides a check independent of that reuse.
Such checks are useful in unresolved limits, where large terms can cancel.

The common high-throughput path is different.  Once process output has been
generated, double-precision evaluation runs through \rusticol{} and the stored
evaluator without loading or calling \symbolica{}.  The default just-in-time
(JIT) evaluator uses the \symjit{} runtime linked
into \rusticol{};
compatible C++ and assembly output loads its own native library.

\symbolica{} generally requires an academic license for multi-core execution. However,
\pyac{} has a \emph{library license}, which allows all of its users to use
\pyac{} in parallel without requiring a \symbolica{} license of their own. All threads created
by \pyac{} are automatically unlocked by calling \texttt{set\_library\_key} upon module loading. A library license is active as long as the developers have an active \symbolica{} license.

\subsection{Numerical kernels with \symjit{}}
\label{sec:symjit}
\symjit{}~\cite{SymJITSoftware} is an optimising just-in-time (JIT) code generator that is the
default backend for \symbolica{}. The \symbolica{}/\symjit{} stack acts as a compiler,
where \symbolica{} supplies optimised scalar formulae and \symjit{} turns them into efficient 
double-precision code.

\symjit{} is specifically designed to process the type of expressions generated by \pyac{}. In the \emph{recurrence} generation mode of \pyac{}, these expressions are limited to those of the individual recursion kernels. However, in the \emph{compiled} generation mode, these
expressions are typically very large, have no loops and only a few forward conditional branches, and are dominated by complex-valued arithmetic operations. A general-purpose compiler backend, such as LLVM, struggles to process these monolithic functions because of $\mathcal{O}(n^2)$ behaviour, where $n$ is the number of operations. On the other hand, \symjit{} generates optimised code with a fraction of the compilation/generation time because it is domain-specific and can focus on the relevant optimisations with a generation time of $\mathcal{O}(n \log(n))$.

\symjit{} applies a multi-level optimisation pipeline to the scalar formulae supplied by \symbolica{}. The result is a linear machine-independent intermediate representation (IR). The IR is portable and can be saved and restored through \symbolica{}. On the target machine, the IR is converted to machine code. At this stage, \symjit{} detects the CPU features and chooses the optimal instructions for the available architecture. Specifically, \symjit{} auto-vectorises to utilise the single-instruction, multiple-data (SIMD) capabilities of the underlying CPU.

For large compiled binaries (kernels), the execution time becomes highly dependent on instruction 
cache misses. The execution time depends non-linearly on the binary size, with clear evidence of a ``phase transition'' near the L3-cache limit. Managing this problem requires close cooperation between \pyac{} and \symjit{}. \symjit{} exposes multiple tuneable switches and optimisation levels, which, among other factors, control the trade-off between speed and size. \pyac{} generates both a small local current kernel, called repeatedly by the recurrence, and a much larger combined evaluator covering a whole stage of a compiled process. Supporting both sizes is important: the former relieves cache pressure, while the latter permits optimisation across a larger part of a particular amplitude.

\rusticol{} arranges the inputs and outputs of these kernels in a shared numerical workspace.  
It can chain them without repeatedly copying intermediate currents, and evaluates independent 
phase-space points in SIMD batches.  The same model rules and formula-to-kernel machinery 
consequently support both a compact recurrence and a larger compiled evaluator, while the 
choice between them remains a performance decision rather than a physics decision.

\subsection{Deriving current kernels with \spenso{}}
\label{sec:spenso}

The numerical kernels described in the preceding subsections require
scalar expressions for the components of the off-shell currents. In
the recurrence introduced above, these expressions are implicit in
the vertex rules and propagators.
\spenso{} \cite{SpensoSoftware} provides the tensor contractions needed to
obtain them. During model processing, \pyac{} selects an output leg
of a UFO Lorentz rule and supplies symbolic currents for the remaining
legs. \spenso{} sums over their indices in a chosen basis, leaving one
scalar expression for each component of the output current. These are
the expressions passed to \symbolica{} for optimisation and compilation.

To illustrate this step, return to the $d\bar d\to Zgg$ example and
consider the last current in eq.~\eqref{eq:zgg-currents},

\[
Q(2,4)=P_d V_{dgd}[Q(2),G(4)].
\]

Here $Q(2)=\bar u_{h_2}(q_2)$ and $G(4)$ are the one-leg currents
introduced there. We use \spenso{}'s tensor API to derive the components
of their quark--gluon join and then apply the propagator $P_d$, keeping
the input components symbolic until numerical evaluation.

For this example, take a massless quark and choose the colour-stripped
vertex normalisation $V^\mu=i\gamma^\mu$ in four-component notation,
with the strong coupling omitted as in eq.~\eqref{eq:zgg-currents}. We use the Weyl gamma
matrices supplied by \spenso{}'s high-energy-physics tensor library and
the metric $g=\mathrm{diag}(1,-1,-1,-1)$. With
$q_{24}=q_2+q_4$ and $s_{24}=q_{24}^2$, the current is then

\[
Q(2,4)_c = \frac{i}{s_{24}}\sum_b
\left[i\sum_{a,\mu}Q(2)_a(\gamma^\mu)_{ab}G(4)_\mu\right]
(\not q_{24})_{bc}.
\]

Here $a,b,c$ are spinor indices and $\mu$ is a Lorentz index, each
taking four values, with $\not q_{24}=\gamma^\nu q_{24,\nu}$.
The bracket is $V_{dgd}[Q(2),G(4)]$; since $Q(2)$ is a row spinor,
the propagator $i\not q_{24}/s_{24}$ acts on its right. Summing the
repeated indices therefore leaves four expressions labelled by $c$.

To express this contraction in code, we first specify the spaces to
which the indices belong: \texttt{Rep.bis(4)} for the four spinor components
and \texttt{Rep.mink(4)} for Minkowski vectors. The latter includes the metric
used in contractions, so vector components can be supplied
contravariantly. With these representations, \texttt{TN.vector} declares
a tensor carrying one index. The arguments \texttt{2}, \texttt{4} and \texttt{2, 4} retain
the leg labels of $Q(2)$, $G(4)$ and $q_{24}$ in their names; tensor
indices are assigned when the contractions are formed below. You can interactively run these examples in this \href{https://symbolica.io/gallery/notebooks/pyamplicol_spenso_current_kernel.html}{\texttt{Python} notebook} hosted on the \symbolica{} gallery.
Each listing is followed by its rendered output.

\begin{notebookcell}{Symbolic input currents and momentum.}{lst:spenso-inputs}
\begin{lstlisting}[style=pythoncode]
from symbolica import Symbol
from symbolica.community.tensor import Representation as Rep
from symbolica.community.tensor import TensorExpression as T
from symbolica.community.tensor import TensorName as TN

spinor, vector = Rep.bis(4), Rep.mink(4)
Q2 = TN.vector("Q")(2, spinor)
G4 = TN.vector("G")(4, vector)
q24 = TN.vector("q")(2, 4, vector)

Q2, G4, q24
\end{lstlisting}
\tcblower
{\small\sffamily\color{codecomment} Output}\par\smallskip
\centering
\includegraphics[width=28.305pt]{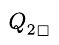}\qquad
\includegraphics[width=28.250pt]{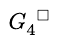}\qquad
\includegraphics[width=29.881pt]{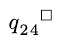}
\end{notebookcell}

The first contraction forms the bracket in the component expression.
\spenso{} orders the indices of a gamma matrix as \texttt{(row, column, Lorentz)}.
Assigning the same label to compatible indices contracts them, so
\texttt{Q2(1)} joins the row index and \texttt{G4(2)} the Lorentz index. The placeholder
\texttt{AUTO}, imported as \texttt{\_}, leaves the remaining spinor index unresolved
so it can be contracted with the propagator in the next step.

\begin{notebookcell}{The quark--gluon vertex contraction.}{lst:spenso-vertex}
\begin{lstlisting}[style=pythoncode]
from symbolica.community.tensor import AUTO as _

gamma = T.dirac_gamma(4)
i = Symbol.I
join = i * gamma(1, _, 2) * Q2(1) * G4(2)
join
\end{lstlisting}
\tcblower
{\small\sffamily\color{codecomment} Output}\par\smallskip
\centering
\includegraphics[width=72.106pt]{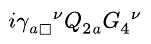}
\end{notebookcell}

The same gamma matrix supplies the propagator numerator when its
Lorentz index is contracted with $q_{24}$, while \texttt{q24 * q24} gives
the denominator $s_{24}$. Here the label \texttt{3} is used for both the
contracted Lorentz index and the open output spinor index. Since they
belong to different spaces, these indices remain distinct. The first
spinor index is left unresolved with \texttt{\_}; labelling the output index
explicitly makes the subsequent contraction unambiguous.

\begin{notebookcell}{The massless quark propagator.}{lst:spenso-propagator}
\begin{lstlisting}[style=pythoncode]
from symbolica.community.tensor import AUTO as _

s24 = q24 * q24
propagator = i * gamma(_, 3, 3) * q24(3) / s24
propagator
\end{lstlisting}
\tcblower
{\small\sffamily\color{codecomment} Output}\par\smallskip
\centering
\includegraphics[width=56.471pt]{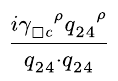}
\end{notebookcell}

Multiplication now contracts the two compatible unresolved spinor indices,
completing the expression for $Q(2,4)$ with only its output index open.
This current has a single vertex contribution. For the three-leg
currents following eq.~\eqref{eq:zgg-currents}, the two contributions would first
be added and the propagator applied to their sum.

\begin{notebookcell}{Applying the propagator to the vertex contribution.}{lst:spenso-current}
\begin{lstlisting}[style=pythoncode]
current = join * propagator
current
\end{lstlisting}
\tcblower
{\small\sffamily\color{codecomment} Output}\par\smallskip
\centering
\includegraphics[width=116.171pt]{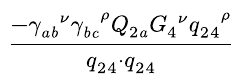}
\end{notebookcell}

Up to this point, \texttt{current} records the tensors and their contractions.
Parsing it with \texttt{to\_network()} turns this expression into an executable
tensor network, using the built-in tensor library for the gamma matrices
and symbolic components for the named inputs. The network retains the
contractions and the scalar operations needed for the propagator
denominator; these are evaluated in the next step.

\begin{notebookcell}{Parsing the current into a tensor network.}{lst:spenso-network}
\begin{lstlisting}[style=pythoncode]
network = current.to_network()
network
\end{lstlisting}
\tcblower
{\small\sffamily\color{codecomment} Output}\par\smallskip
\centering
\includegraphics[width=201.671pt]{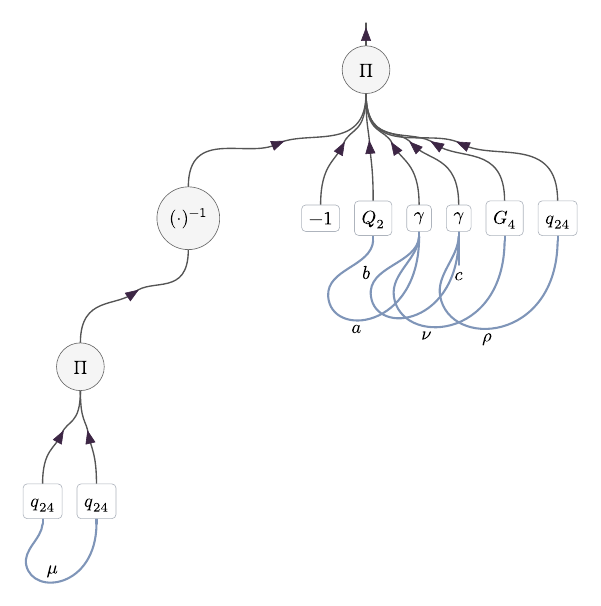}
\end{notebookcell}

Executing the network carries out the component sums and scalar
operations, leaving the input components symbolic. From the completed
network, \texttt{result\_tensor()} extracts the four-component current $Q(2,4)$.
We display its first component, \texttt{Q24[0]}, below; the complete tensor is
retained for numerical evaluation. Numeric superscripts in the output
label entries in the chosen basis. Line breaks have been added to fit
the expression on the page.

\begin{notebookcell}{Executing the network and displaying the first current component.}{lst:spenso-components}
\begin{lstlisting}[style=pythoncode]
network.execute()
Q24 = network.result_tensor()
Q24[0]
\end{lstlisting}
\tcblower
{\small\sffamily\color{codecomment} Output (line-wrapped)}\par\smallskip
\centering
\includegraphics[width=354.167pt]{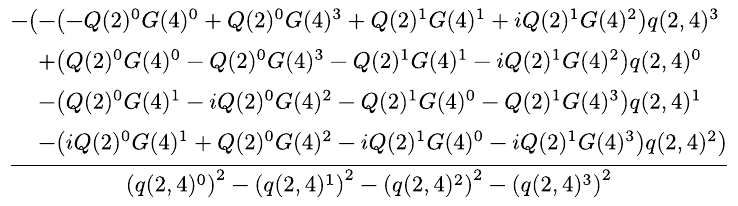}
\end{notebookcell}

The full current can now be evaluated through the \symbolica{} interface
introduced above. The tensor's \texttt{evaluator} method accepts the same
options as \texttt{Expression.evaluator} and optimises all four components
together. We use \texttt{components()} to list its inputs in the order
$Q(2)$, $G(4)$, $q_{24}$; each numerical input row then produces a
four-component tensor. As in the earlier scalar example, \symjit{}
compiles the evaluator on its first numerical use.

\begin{notebookcell}{Numerical evaluation of the current.}{lst:spenso-evaluation}
\begin{lstlisting}[style=pythoncode]
parameters = [*Q2.components(), *G4.components(), *q24.components()]
evaluator = Q24.evaluator(parameters)

quark = [2, 1 - 1j, 0, 0]
gluon = [0, 0, 4 / 5, -3 / 5]
momentum = [8, 2, 5, 5]
value = evaluator.evaluate_complex([[*quark, *gluon, *momentum]])[0]
value
\end{lstlisting}
\tcblower
{\small\sffamily\color{codecomment} Output}\par\smallskip
\centering
\includegraphics[width=300.564pt]{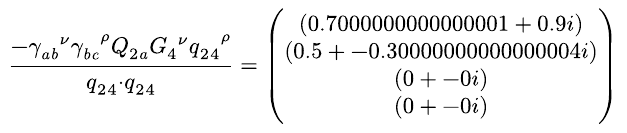}
\end{notebookcell}

The numerical inputs above correspond to lightlike momenta
$q_2=(3,2,2,1)$ and $q_4=(5,0,3,4)$, with
$Q(2)=(2,1-i,0,0)$ and $G(4)=(0,0,4/5,-3/5)$. The spinor satisfies
$Q(2)\not q_2=0$, while the linearly polarised gluon obeys
$q_4\cdot G(4)=0$ and $G(4)^2=-1$. The spinor's overall normalisation
is arbitrary. Their sum gives $q_{24}=(8,2,5,5)$ and $s_{24}=10$, so

\[
Q(2,4)=\left(\frac{7+9i}{10},\frac{5-3i}{10},0,0\right).
\]

The two vanishing components reflect the chirality of the input
spinor, which is preserved by the vertex and massless propagator
together. Only the inputs to this local join are needed for the
evaluation.

Returning to the recurrence, $Q(2,4)$ supplies both
$V_{dZd}[Q(2,4),Z(3)]$ in $Q(2,3,4)$ and
$V_{dgd}[Q(2,4),G(5)]$ in $Q(2,4,5)$. The shared-current construction
ensures that its value is computed once for these two uses. Each later
join has its own component rule derived in the same way, until the
final sum is contracted with the closing-leg spinor $v_{h_1}(q_1)$
without a propagator, as in eq.~\eqref{eq:zgg-closure}.

Deriving numerical routines from interaction tensors is also the purpose
of \aloha{} \cite{deAquino:2011ub} and the UFO extension of \comix{}
\cite{Hoeche:2014kca}. \spenso{} provides a generic contraction algorithm in
which the index spaces and tensor entries are supplied separately.
Thus the example can be expressed in a different Dirac basis by
supplying the corresponding matrices and current components, while
retaining the same contractions. The tensor library also provides
$\gamma^5$, chiral projectors and fundamental or adjoint $SU(3)$
representations; additional tensors can be defined through their
symbolic or numerical entries. Before evaluating components, the
symbolic tensor-algebra library \idenso{} can apply metric, Dirac and
colour identities through \texttt{TensorExpression.simplify\_algebra}.

The open indices need not describe a single spinor or vector. In the
momentum-independent four-point construction of section~\ref{sec:general-contact-decomposition}, contracting two inputs leaves the intermediate
tensor $X_{\rho\sigma}$ with two Lorentz indices. \spenso{} obtains its
16 component expressions in the same way as the four spinor components
of $Q(2,4)$. \symbolica{} then identifies exact zeros and components related by
equality or a sign, giving the one stored representative for the
metric-pair structure and the six for the antisymmetric four-gluon
structure discussed there. In the four-gluon case, only the six entries
with $\rho<\sigma$ need to be stored: the others follow from
$X_{\rho\rho}=0$ and $X_{\sigma\rho}=-X_{\rho\sigma}$. \pyac{}
records these relations so that the second join can contract with the
remaining input current using the six stored entries, with exactly
the same result as using all 16.

For the supported higher-point colour-singlet interactions where no
compact intermediate current is constructed, \pyac{} retains the
input currents and momenta until the final join. \spenso{} derives the
component expressions for that join by contracting the original UFO
interaction with all its inputs. Thus \pyac{} specifies which
intermediate currents to build, while \spenso{} supplies the expressions
needed to evaluate the chosen joins. Section~\ref{sec:general-contact-decomposition} describes these constructions and their validity checks;
table~\ref{tab:ufo-coverage} lists the interactions currently supported.

\section{Results}
\label{sec:results}

Numerical validation covered the built-in Standard Model with the HEFT extension, and the same model
loaded from its UFO and JSON forms over a broad set of QCD and electroweak
processes.  The checks included leading-, next-to-leading- and full-colour
evaluation and selected and summed helicities.  Where an independently generated \ac{} result existed, it was
compared at the same phase-space point.  Among 497 comparisons with
independently generated \ac{} results, the largest relative difference was
$1.1\times10^{-11}$.  Across a further 497 comparisons of compiled or eager
evaluation with recurrence, the maximum was $3\times10^{-13}$.  In 1238
checks, matrix elements returned by summed colour-and-helicity requests agreed
with the explicit sum of their resolved contributions to
$3\times10^{-13}$ or better at regular phase-space points.  The
benchmark definitions and full tables are given in
appendix~\ref{app:benchmarks}; the independent full-colour comparison in
appendix~\ref{app:mg-comparison} provides a complementary check.

\subsection{\texorpdfstring{$Z$}{Z} production with additional gluons}
\label{sec:z-results}

We use $d\bar d\to Z+(n-1)g$ as a representative benchmark
for a colourless boson accompanied by QCD radiation.
Table~\ref{tab:z-progression} shows the multiplicity dependence for
$d\bar d\to Z+(n-1)g$, where $n$ is the number of final-state particles.  The
LC comparisons evaluate one physical colour flow, either summing all external
helicities or selecting one non-zero helicity configuration, through $n=9$.
The same retained \pyac{} process outputs are used for both requests and
remain valid for every helicity: the restriction is imposed only at runtime.
The full-colour workload stops at $n=7$;
it sums the helicities and contracts all
colour interferences.  Structurally vanishing helicities contribute zero and
do not require a numerical kernel.  The tabulated \pyac{} times use
recurrence JIT-O2 throughout. Full-colour entries select the fastest tested,
validated colour-contraction mode;
the superscript identifies adaptive-adjoint FFT, trace FFT or direct
contraction. For this quark process the adaptive option retains the natural
fundamental-chain basis, rather than introducing a gluonic DDM basis.

For a single helicity, we choose the alternating assignment
\begin{equation}
  (h_d,h_{\bar d},\lambda_Z,\lambda_{g_1},\ldots)
  =(-,+,-,+,-,\ldots),
  \label{eq:z-benchmark-helicity}
\end{equation}
in external-particle order $(d,\bar d,Z,g_1,\ldots)$, where $h$ denotes a fermion helicity, $\lambda$ a vector-boson helicity, the subscript names the external particle, the gluons $g_1,\ldots,g_{n-1}$ are numbered in external-state order, and the signs denote
fermion helicities $\pm\tfrac12$ and transverse-vector helicities $\pm1$.
Every selected matrix element is non-zero at the benchmark point.
For $n\geq4$, the corresponding massless helicity assignment has at least
three helicities of each sign after crossing to the all-outgoing convention of eq.~\eqref{eq:all-outgoing},
so it is neither maximally helicity-violating (MHV) nor its parity conjugate.
The same prescription fixes
the lower multiplicities, where such a non-MHV sector is unavailable.
The selected colour ordering, listing the colour-adjacent legs along the open quark line from the antiquark to the quark as in the colour words of eq.~\eqref{eq:zgg-words}, is $(\bar d,g_1,\ldots,g_{n-1},d)$, with
the gluons ordered as in the external state.

The codes evaluate the same regular phase-space point, with
the strong coupling $\alpha_s=0.118$, the inverse fine-structure constant $\alpha^{-1}=132.507$, the $Z$-boson mass $M_Z=91.188\,\mathrm{GeV}$ and
the Fermi constant $G_F=1.16639\times10^{-5}\,\mathrm{GeV}^{-2}$.
The centre-of-mass energy is $1\,\mathrm{TeV}$ for $n\geq2$ and $M_Z$ for
the on-shell $2\to1$ process $d\bar d\to Z$, the $n=1$ entry.  We include the initial-state averages and
identical-particle factors.  \mgamc{} provides an additional
independent full-colour comparison. The timing protocol and reference release
are specified in the caption and appendix~\ref{app:benchmark-reproducibility}.

\begin{table}[htbp]
  \centering
  \caption{Warmed double-precision evaluation times for
  $d\bar d\to Z+(n-1)g$ on one Apple M3 Pro CPU core.
  A sample is one phase-space point.
  The upper panel compares one physical LC flow with selected or summed
  helicities; the lower panel includes all colour interferences and sums
  the helicities. All times include amplitude squaring.
  The selected helicity is $(h_d,h_{\bar d},\lambda_Z,\lambda_{g_1},\ldots)
  =(-,+,-,+,-,\ldots)$ in external-particle order, with fermion helicities
  $\pm\tfrac12$ and transverse-vector helicities $\pm1$.
  The retained \pyac{} process outputs support all helicities in both LC
  measurements. The \pyac{} time columns use recurrence JIT-O2 throughout.
  Superscripts A, T and D identify adaptive-adjoint FFT, trace FFT and
  direct colour contraction. Marked entries use the fastest validated
  mode tested for that process; A retains the fundamental-chain basis
  for this quark process.
  Coloured ratios give the tabulated \pyac{}/\ac{} time, green below one
  and orange above one. All ratios are calculated before rounding to three
  significant figures.
  The reference interfaces and timing protocol are specified in
  appendix~\ref{app:benchmark-reproducibility}.}
  \label{tab:z-progression}
  \small
  \renewcommand{\arraystretch}{1.1}
  \setlength{\tabcolsep}{2pt}
  \sisetup{table-text-alignment=center}
  \begin{tabular*}{\linewidth}{@{\extracolsep{\fill}}c
      S[table-format=1.4] S[table-format=1.4]
      c
      S[table-format=3.4] S[table-format=3.3]
      c@{}}
    \toprule
    \multicolumn{7}{c}{$d\bar d\to Z+(n-1)g$} \\
    \addlinespace[0.5ex]
    \multicolumn{7}{c}{Single LC flow [\si{\micro\second}/sample]} \\
    & \multicolumn{3}{c}{One helicity} & \multicolumn{3}{c}{Helicity sum} \\
    \cmidrule(lr){2-4}\cmidrule(lr){5-7}
    $n$ & {\ac{}} & {\pyac{}} & {ratio}
        & {\ac{}} & {\pyac{}} & {ratio} \\
    \midrule
    1 & 0.0494 & 0.0315 & \textcolor{green!50!black}{\tablenum[table-format=1.3]{0.637}} & 0.0285 & 0.109 & \textcolor{orange!85!black}{\tablenum[table-format=1.3]{3.83}} \\
    2 & 0.117 & 0.0625 & \textcolor{green!50!black}{\tablenum[table-format=1.3]{0.533}} & 0.164 & 0.334 & \textcolor{orange!85!black}{\tablenum[table-format=1.3]{2.04}} \\
    3 & 0.220 & 0.121 & \textcolor{green!50!black}{\tablenum[table-format=1.3]{0.550}} & 0.498 & 0.824 & \textcolor{orange!85!black}{\tablenum[table-format=1.3]{1.65}} \\
    4 & 0.436 & 0.222 & \textcolor{green!50!black}{\tablenum[table-format=1.3]{0.510}} & 1.57 & 2.32 & \textcolor{orange!85!black}{\tablenum[table-format=1.3]{1.48}} \\
    5 & 0.790 & 0.360 & \textcolor{green!50!black}{\tablenum[table-format=1.3]{0.456}} & 4.84 & 5.39 & \textcolor{orange!85!black}{\tablenum[table-format=1.3]{1.11}} \\
    6 & 1.33 & 0.602 & \textcolor{green!50!black}{\tablenum[table-format=1.3]{0.454}} & 14.1 & 13.4 & \textcolor{green!50!black}{\tablenum[table-format=1.3]{0.951}} \\
    7 & 2.08 & 0.993 & \textcolor{green!50!black}{\tablenum[table-format=1.3]{0.477}} & 41.6 & 35.4 & \textcolor{green!50!black}{\tablenum[table-format=1.3]{0.851}} \\
    8 & 3.10 & 1.50 & \textcolor{green!50!black}{\tablenum[table-format=1.3]{0.483}} & 106 & 87.3 & \textcolor{green!50!black}{\tablenum[table-format=1.3]{0.824}} \\
    9 & 4.62 & 2.28 & \textcolor{green!50!black}{\tablenum[table-format=1.3]{0.493}} & 272 & 218 & \textcolor{green!50!black}{\tablenum[table-format=1.3]{0.801}} \\
    \bottomrule
  \end{tabular*}
  \par\vspace{1.2ex}
  \begin{tabular*}{\linewidth}{@{\extracolsep{\fill}}c
      S[table-format=6.3] S[table-format=6.3]
      S[table-format=5.3]
      c@{}}
    \toprule
    \multicolumn{5}{c}{Full colour, helicity sum [\si{\micro\second}/sample]} \\
    \addlinespace[0.4ex]
    $n$ & {\mgfive{}} & {\ac{}} & {\pyac{}} & {ratio} \\
    \midrule
    1 & 0.302 & 0.102 & \multicolumn{1}{c}{\tablenum[table-format=5.3]{0.110}\textsuperscript{T}} & \textcolor{orange!85!black}{\tablenum[table-format=1.3]{1.08}} \\
    2 & 1.37 & 0.279 & \multicolumn{1}{c}{\tablenum[table-format=5.3]{0.318}\textsuperscript{A}} & \textcolor{orange!85!black}{\tablenum[table-format=1.3]{1.14}} \\
    3 & 7.77 & 1.61 & \multicolumn{1}{c}{\tablenum[table-format=5.3]{1.33}\textsuperscript{D}} & \textcolor{green!50!black}{\tablenum[table-format=1.3]{0.827}} \\
    4 & 66.0 & 14.7 & \multicolumn{1}{c}{\tablenum[table-format=5.3]{10.6}\textsuperscript{T}} & \textcolor{green!50!black}{\tablenum[table-format=1.3]{0.724}} \\
    5 & 888 & 208 & \multicolumn{1}{c}{\tablenum[table-format=5.3]{116}\textsuperscript{D}} & \textcolor{green!50!black}{\tablenum[table-format=1.3]{0.558}} \\
    6 & 21200 & 4360 & \multicolumn{1}{c}{\tablenum[table-format=5.3]{2120}\textsuperscript{T}} & \textcolor{green!50!black}{\tablenum[table-format=1.3]{0.487}} \\
    7 & 920000 & 173000 & \multicolumn{1}{c}{\tablenum[table-format=5.3]{43800}\textsuperscript{T}} & \textcolor{green!50!black}{\tablenum[table-format=1.3]{0.253}} \\
    \bottomrule
  \end{tabular*}
\end{table}

The independent matrix-element evaluations agree within a relative difference of $\num{4e-11}$.

For the single-helicity comparison, both the specified configuration and
its global helicity reversal, in which the sign of every external helicity in eq.~\eqref{eq:z-benchmark-helicity} is flipped, agree between the codes within a relative
difference of $3\times10^{-12}$.

For a selected helicity and flow, both \pyac{} backends retain only the
currents needed by that request, including their shared subcurrents.
The same process output therefore supports helicity sampling and complete
helicity sums, without evaluating unused configurations on each selected
call.

Process setup also has a different role in the two programs.  A compiled
\pyac{} process includes process-specific formula construction and numerical
kernel preparation, whereas \ac{} starts from a specialised implementation.
The setup cost is amortised when the matrix element is evaluated many times,
as in phase-space integration and event generation. Setup time and memory are
kept separate from warmed evaluation in the accompanying benchmark data.
We also stress that \ac{} was designed and optimised for evaluating a
single colour ordering of helicity-summed matrix elements, for which its performance is closest to that of \pyac{}.

\FloatBarrier
\subsection{Higgs production in the gluon effective theory}
\label{sec:heft-results}

Table~\ref{tab:heft-progression} applies the same three workloads to
$gg\to H+(n-1)g$ in the Higgs--gluon effective theory, retaining one
insertion of the effective Higgs interaction. This process family probes
the treatment of higher-point interactions alongside the ordinary QCD
recursion. The two single-flow LC columns report \pyac{} alone;
for the full-colour helicity sum, the independent comparison and timing
ratio use \mgfive{} through $n=7$. Superscripted full-colour entries select
the fastest validated mode tested. The adaptive option uses
the certified DDM basis for this single-insertion HEFT family. Local adjoint
generation at $n=7$ exceeded the 30\,GB memory limit, so that process was
generated on a larger-memory cluster and timed locally under the same runtime
guard. The squared matrix elements agree to a relative difference of
$3.3\times10^{-15}$. The full-colour helicity sum takes
$0.664\,\mathrm{s}$, compared with $24.4\,\mathrm{s}$ for \mgfive{}, a
speedup of a factor $36.8$.
Generation remains generic in helicity and colour flow, and only the
runtime request selects the desired components.
The selected-helicity prescription is given with the table.
Large \mgfive{} matrix routines are divided into helper subroutines in
source files of at most approximately 1\,MB before compilation. The files
are compiled separately, preserving the generated algebra;
details are given in appendix~\ref{app:benchmark-reproducibility}.

The full-colour, helicity-summed $n=7$ entry in
Table~\ref{tab:heft-progression} probes a resource-intensive regime,
with binary process outputs reaching the gigabyte
scale. Performance then depends strongly on how these large data sets are
stored and reused in memory, so the observed timings cannot be understood
from algorithmic scaling alone. With the reduced DDM output, the local
profiling process peaks at $5.23\,\mathrm{GB}$, well below the runtime guard.

\begin{table}[htbp]
  \centering
  \caption{Warmed double-precision evaluation times for
  $gg\to H+(n-1)g$ in the Higgs--gluon effective theory, with $n$
  final-state particles. A sample is one phase-space point; all times
  include amplitude squaring. The LC columns give \pyac{} times for one
  physical colour flow, with a selected helicity or the full helicity sum.
  Process outputs retain support for all helicities and LC flows;
  all \pyac{} results use recurrence JIT-O2.
  The seed alternates gluon helicities $(-,+,-,+,\ldots)$ in external
  order $(g_1,g_2,H,g_3,\ldots)$, with zero Higgs helicity. If necessary,
  the fewest gluon helicities are changed to obtain a nonzero value;
  $\lambda_i$ denotes the helicity of external leg $i$.
  Full-colour results sum helicities and include all colour interferences.
  Their coloured ratio is $t_{\rm pyAC}/t_{\rm MG5}$, green below one
  and orange above one, calculated before rounding to three significant figures.
  Full-colour entries beyond $n=7$ are outside this comparison and are blank.
  Changes to the alternating seed are $n=1$: $\lambda_{2}=-1$.
  Superscripts A, T and D identify adaptive-adjoint FFT, trace FFT and
  direct colour contraction. Marked entries use the fastest validated
  mode tested for that process; A uses the certified DDM basis here.
  Full-colour \pyac{} at $n=7$ retains the original one-point profiling protocol under a 30\,GB runtime guard; the other \pyac{} entries use batches of 128 points.
  The \mgfive{} time at $n=7$ is one measured scalar evaluation after one warm-up call.
  Timing and model settings, including the subdivision of large
  \mgfive{} routines before compilation, are described in
  appendix~\ref{app:benchmark-reproducibility}.}
  \label{tab:heft-progression}
  \small
  \renewcommand{\arraystretch}{1.12}
  \setlength{\tabcolsep}{3pt}
  \sisetup{table-text-alignment=center}
  \begin{tabular*}{\linewidth}{@{\extracolsep{\fill}}c
      S[table-format=2.4] S[table-format=4.3] S[table-format=8.4] S[table-format=6.3]
      c@{}}
    \toprule
    \multicolumn{6}{c}{$gg\to H+(n-1)g$} \\
    \addlinespace[0.5ex]
    \multicolumn{6}{c}{Evaluation time [\si{\micro\second}/sample]} \\
    & \multicolumn{2}{c}{Single LC flow: \pyac{}}
      & \multicolumn{3}{c}{Full colour, helicity sum} \\
    \cmidrule(lr){2-3}\cmidrule(lr){4-6}
    $n$ & {One helicity} & {Helicity sum} & {\mgfive{}} & {\pyac{}} & {ratio} \\
    \midrule
    1 & 0.0689 & 0.141 & 0.0764 & \multicolumn{1}{c}{\tablenum[table-format=6.3]{0.137}\textsuperscript{T}} & \textcolor{orange!85!black}{\tablenum[table-format=1.4]{1.79}} \\
    2 & 0.910 & 3.36 & 1.21 & \multicolumn{1}{c}{\tablenum[table-format=6.3]{3.26}\textsuperscript{A}} & \textcolor{orange!85!black}{\tablenum[table-format=1.4]{2.70}} \\
    3 & 1.48 & 6.63 & 14.7 & \multicolumn{1}{c}{\tablenum[table-format=6.3]{9.60}\textsuperscript{A}} & \textcolor{green!50!black}{\tablenum[table-format=1.4]{0.653}} \\
    4 & 2.77 & 20.7 & 257 & \multicolumn{1}{c}{\tablenum[table-format=6.3]{109}\textsuperscript{A}} & \textcolor{green!50!black}{\tablenum[table-format=1.4]{0.424}} \\
    5 & 5.14 & 62.9 & 7610 & \multicolumn{1}{c}{\tablenum[table-format=6.3]{1400}\textsuperscript{A}} & \textcolor{green!50!black}{\tablenum[table-format=1.4]{0.184}} \\
    6 & 9.12 & 179 & 588000 & \multicolumn{1}{c}{\tablenum[table-format=6.3]{22000}\textsuperscript{A}} & \textcolor{green!50!black}{\tablenum[table-format=1.4]{0.0373}} \\
    7 & 16.6 & 495 & 24400000 & \multicolumn{1}{c}{\tablenum[table-format=6.3]{664000}\textsuperscript{A}} & \textcolor{green!50!black}{\tablenum[table-format=1.4]{0.0272}} \\
    8 & 27.3 & 1440 &  &  &  \\
    9 & 37.4 & 4090 &  &  &  \\
    \bottomrule
  \end{tabular*}
\end{table}

\subsection{Choosing the colour basis and summation method}
\label{sec:colour-mode-comparison}
\label{sec:fft-results}

Both colour transforms are validated against direct contraction with the
full overlap matrix, including its off-diagonal interferences. The timings
compare complete matrix-element evaluations, including the construction of
partial amplitudes, rather than the colour contraction alone. At fixed
basis the FFT leaves the amplitude recursion and helicity coverage unchanged;
the DDM reduction can additionally reduce the number of ordered amplitudes.
For a complementary comparison of direct, multiplet and FFT methods at
fixed gluon helicities, see ref.~\cite{ColourFFT:2026}.

The benchmarks show that the choice of colour basis can make a substantial
difference to the evaluation time. For $gg\to5g$ in
Table~\ref{tab:mg-comparison}, the full-colour helicity sum takes
$1.43\,\mathrm{ms}$ with adjoint FFT, compared with $9.30\,\mathrm{ms}$
with trace FFT, a speedup of $6.50$. The smaller adjoint basis introduced in
section~\ref{sec:colour-sums} requires fewer partial amplitudes and less work
to combine them. Both calculations already use an FFT, so this gain comes
from changing the basis, not from switching on the transform. These times
include the complete matrix-element evaluation, with all helicities summed.

A similar benefit is seen for $gg\to H+4g$ in the Higgs effective theory,
the $n=5$ row of Table~\ref{tab:heft-progression}. Adjoint FFT takes
$1.40\,\mathrm{ms}$, compared with $6.04\,\mathrm{ms}$ for trace FFT,
a factor of $4.30$. Whether the FFT alone helps is less clear at this
multiplicity. Repeated measurements using the same saved process outputs
gave average times of $7.04\,\mathrm{ms}$ for trace FFT and
$6.24\,\mathrm{ms}$ for direct colour summation, reversing their order in
the first comparison. The colour sum itself was slightly slower with the
FFT, while the rest of the evaluation varied between runs despite doing
the same work. We therefore find no reproducible advantage from the
transform alone here; the clear improvement comes from the smaller adjoint
basis. All three methods agree on the matrix element, and the measurements
and their uncertainties are retained with the table data.

For the quark processes studied here, the adaptive and trace options use
the same colour basis, suited to the open quark lines. There is therefore
no corresponding reduction in the number of partial amplitudes. The direct
method also already exploits some of the symmetries under gluon exchanges,
through the Walsh--Hadamard transform described in
section~\ref{sec:colour-sums}. The FFT must improve on this existing
simplification while paying the extra cost of rearranging and transforming
the amplitudes. With several quark lines, the gluons are distributed among
them, so the number of gluons alone is not a reliable guide to the benefit.

For $d\bar d\to Z+4g$, direct summation is faster and is selected in
Table~\ref{tab:z-progression}. With six final-state gluons, the balance
reverses: trace FFT takes $43.8\,\mathrm{ms}$ compared with
$54.6\,\mathrm{ms}$ for direct summation, a speedup of $1.25$.
In this case the colour sum itself becomes faster, taking $12.7$ rather
than $20.1\,\mathrm{ms}$. The adaptive option uses the same basis and gives
an equivalent timing within uncertainty. These examples show why the tables
report the fastest tested method for each process: neither the basis choice
nor the summation method gives the same benefit in every calculation.

\FloatBarrier
\section{Conclusion}
\label{sec:conclusion}

We have presented \pyac, a generator of fast colour-ordered tree-level matrix
elements.  It retains the central physics idea of \ac---direct access to
individual leading-colour contributions---and generalises the current
construction to a broader class of model rules and processes with more open
quark lines.
Its capabilities include the supported subset of UFO model input,
LC, NLC and full colour, selected or summed helicities,
batched evaluation and precision-controlled Python calculations.  The resulting
ordinary process output is accessible through Python, C, C++, Fortran or Rust.

The central gain comes from addressing two sources of multiplicity growth.
Shared-current recursion avoids the factorial diagram expansion and gives
polynomial cost for one gluon ordering at fixed helicity.  Exact colour
summation is a separate problem: direct interference of factorially many
orderings has factorial-squared cost.  The symmetric-group FFT~\cite{ColourFFT:2026}
replaces that dense sum by a fast transform and smaller contractions without
approximating the colour algebra.  The number of ordered amplitudes and
helicities still grows, but their direct colour double sum is no longer
required on the complete permutation orbits.

The implementation was validated against the original \ac{} and independent
matrix-element calculations, and its several numerical arrangements were
cross-checked against one another.  In the benchmarks reported here, \pyac{}
reached or exceeded the \ac{} evaluation speed for several full-colour and
higher-multiplicity workloads, while supporting a broader class of model rules.
Its advantage is most visible when the complete colour contraction is required, while a
compact recurrence provides favourable scaling when a very large compiled
expression would be expensive.  The additional gain from colour transforms
depends on the available permutation symmetries and on exact reductions
already used in the direct calculation.

These features target the applications described in the introduction:
high-multiplicity matrix-element samples and merging, matrix-element
corrections in parton showers, and real-radiation contributions in precision
calculations.  Colour-flow and helicity resolution are useful in their own
right in showers.  The correlated Born interface additionally retains the
interferences needed for colour insertions and literal spin projections,
with a generic construction of ordered colour connections at arbitrary
perturbative order and LC, NLC or full colour.  Higher-precision evaluation,
where the symbolic expression is retained,
supports validation and rescue calculations near singular limits.  The local
collinear and soft subtraction studies illustrate how it extends the range in
which a small difference of large matrix elements is resolved.  The supported UFO
route and common language interfaces make the same physics available without
binding an application to one generated source language.

The work also served as an experiment in agentic development.  A proprietary
AI model, guided and repeatedly checked by domain experts, helped produce a
substantial scientific implementation rapidly.  We set the physics requirements
and checked the results against independent calculations.  Releasing the
program, its validation tests and its benchmark tools makes that outcome
available beyond the experiment itself.

\section*{Acknowledgements}
\label{sec:acknowledgements}

We thank the wider \mgamc{} community,
whose tools and model conventions made the independent checks possible.  We
are grateful to Malin Sj\"odahl for discussions of colour bases and Fourier
methods.  V.H. acknowledges support from the Swiss National Science Foundation
under grant PCEFP2\_203335. The other authors thank V.H. for personally covering
the AI subscription costs that supported this work.

\appendix
\section{Benchmarked performance}
\label{app:benchmarks}

This appendix compares one leading-colour flow across several process families
and full-colour results against independent generators.  The measurement
protocol and full tables are given below, with phase-space inputs, unrounded
measurements and resource-limited results supplied in the accompanying data.  Reported ratios are calculated before rounding the
individual times, so they need not equal the quotient of the displayed values.

\subsection{One leading-colour flow}
\label{app:lc-matrix}

Table~\ref{tab:lc-matrix} compares eight process families.
Here $n$ is the number of final-state particles.  Each entry contains
the absolute \pyac{} time and the ratio
\begin{equation}
  R_{\rm LC}=\frac{t_{\rm pyAC}}{t_{\rm AC}},
  \label{eq:appendix-lc-ratio}
\end{equation}
for one physical LC flow and the complete physical helicity sum, where $t_{\rm pyAC}$ and $t_{\rm AC}$ are the \pyac{} and \ac{} evaluation times per phase-space point for the same workload, so that a value below one favours \pyac{}.  The faster
of validated compiled and recurrence evaluation is selected, with the modes
and setup limits specified in the caption.  There and below, JIT-O2 and JIT-O3 denote the just-in-time (JIT) optimisation level, 2 or 3, at which \symjit{} generated the numerical kernels of a process output; level-2 outputs are portable between machines, whereas level-3 code is specific to the machine that generated it.
The shared $Z$-plus-jets entries correspond to the helicity-summed LC block
of Table~\ref{tab:z-progression}, not its single-helicity block.

\begin{table}[H]
  \centering
  \caption{Warmed evaluation time for one selected leading-colour flow and the
  complete physical helicity sum, with $n$ final-state particles.
  Each entry gives the \pyac{} time in black, in $\mu$s/sample, followed by
  the multiplier $R_{\rm LC}=t_{\rm pyAC}/t_{\rm AC}$ in parentheses.
  A sample is one phase-space point.  Times and multipliers have three
  significant figures; multipliers are calculated before rounding, with orange above
  one and green below one.  Superscripts $c$ and $r$ identify the selected
  compiled JIT-O3 or recurrence JIT-O2 measurement.  Recurrence is used when
  compiled generation exceeds the 1\,h or 10\,GB per-worker limit.
  Both implementations include helicity summation and amplitude squaring;
  \ac{} uses single-core CPU time; \pyac{} uses batched wall time.
  The $d\bar d\to Z+(n-1)g$ row shares the recurrence and \ac{} timings
  of Table~\ref{tab:z-progression}, without separate remeasurement.
  }
  \label{tab:lc-matrix}
  \small
  \renewcommand{\arraystretch}{1.12}
  \setlength{\tabcolsep}{3pt}
  \begin{tabularx}{\textwidth}{Y c c c c}
    \toprule
    Process family & $n=4$ & $n=5$ & $n=6$ & $n=7$ \\
    \midrule
    $d\bar d\to Z+(n-1)g$
      & \textcolor{black}{$2.32$}\,\textcolor{orange!85!black}{$({\times} 1.48^{r})$} & \textcolor{black}{$5.39$}\,\textcolor{orange!85!black}{$({\times} 1.11^{r})$} & \textcolor{black}{$13.4$}\,\textcolor{green!50!black}{$({\times} 0.951^{r})$} & \textcolor{black}{$35.4$}\,\textcolor{green!50!black}{$({\times} 0.851^{r})$} \\
    $u\bar d\to e^+\nu_e+(n-2)g$
      & \textcolor{black}{$0.305$}\,\textcolor{orange!85!black}{$({\times} 1.78^{c})$} & \textcolor{black}{$0.866$}\,\textcolor{orange!85!black}{$({\times} 1.35^{c})$} & \textcolor{black}{$2.99$}\,\textcolor{orange!85!black}{$({\times} 1.28^{c})$} & \textcolor{black}{$7.84$}\,\textcolor{orange!85!black}{$({\times} 1.03^{r})$} \\
    $d\bar d\to ZZ+(n-2)g$
      & \textcolor{black}{$2.57$}\,\textcolor{orange!85!black}{$({\times} 1.10^{c})$} & \textcolor{black}{$7.25$}\,\textcolor{orange!85!black}{$({\times} 1.08^{c})$} & \textcolor{black}{$17.2$}\,\textcolor{green!50!black}{$({\times} 0.861^{r})$} & \textcolor{black}{$42.4$}\,\textcolor{green!50!black}{$({\times} 0.830^{r})$} \\
    $gg\to t\bar t+(n-2)g$
      & \textcolor{black}{$3.77$}\,\textcolor{orange!85!black}{$({\times} 1.44^{c})$} & \textcolor{black}{$13.7$}\,\textcolor{orange!85!black}{$({\times} 1.63^{c})$} & \textcolor{black}{$45.6$}\,\textcolor{orange!85!black}{$({\times} 1.87^{r})$} & \textcolor{black}{$112$}\,\textcolor{orange!85!black}{$({\times} 1.74^{r})$} \\
    $gg\to gg+(n-2)g$
      & \textcolor{black}{$4.02$}\,\textcolor{orange!85!black}{$({\times} 1.15^{c})$} & \textcolor{black}{$11.2$}\,\textcolor{orange!85!black}{$({\times} 1.02^{r})$} & \textcolor{black}{$36.6$}\,\textcolor{orange!85!black}{$({\times} 1.05^{r})$} & \textcolor{black}{$75.2$}\,\textcolor{green!50!black}{$({\times} 0.821^{r})$} \\
    $d\bar d\to e^+e^-ZH+(n-4)g$
      & \textcolor{black}{$1.08$}\,\textcolor{orange!85!black}{$({\times} 1.37^{c})$} & \textcolor{black}{$2.49$}\,\textcolor{orange!85!black}{$({\times} 1.42^{c})$} & \textcolor{black}{$4.63$}\,\textcolor{orange!85!black}{$({\times} 1.11^{c})$} & \textcolor{black}{$11.2$}\,\textcolor{orange!85!black}{$({\times} 1.02^{r})$} \\
    $d\bar d\to e^+e^-e^+e^-+(n-4)g$
      & \textcolor{black}{$1.25$}\,\textcolor{green!50!black}{$({\times} 0.855^{c})$} & \textcolor{black}{$2.46$}\,\textcolor{green!50!black}{$({\times} 0.866^{c})$} & \textcolor{black}{$4.52$}\,\textcolor{green!50!black}{$({\times} 0.659^{c})$} & \textcolor{black}{$11.4$}\,\textcolor{green!50!black}{$({\times} 0.629^{r})$} \\
    $d\bar d\to u\bar u s\bar s+(n-4)g$
      & \textcolor{black}{$0.474$}\,\textcolor{orange!85!black}{$({\times} 2.52^{c})$} & \textcolor{black}{$0.980$}\,\textcolor{orange!85!black}{$({\times} 2.20^{c})$} & \textcolor{black}{$1.64$}\,\textcolor{orange!85!black}{$({\times} 1.43^{c})$} & \textcolor{black}{$4.18$}\,\textcolor{orange!85!black}{$({\times} 1.27^{c})$} \\
    \bottomrule
  \end{tabularx}
\end{table}

The relative cost depends on the process family and is not a monotonic
function of the number of final-state particles.

Table~\ref{tab:lc-single-helicity} repeats this comparison for one non-zero
helicity configuration and the same LC flow. The \pyac{} process outputs
remain valid for all helicities and flows; only the runtime selection changes.
The alternating-sign prescription and its exceptions are specified with the
table. This workload is relevant when helicities and colour flows are sampled
in Monte Carlo integration.

\begingroup
\edef\savedLCSingleHelicityTable{\number\value{table}}
\renewcommand{\thetable}{\savedLCSingleHelicityTable\space bis}
\renewcommand{\theHtable}{lc-single-helicity-bis}
\begin{table}[H]
  \centering
  \caption{Warmed evaluation of one leading-colour flow and one nonzero
  helicity configuration, with $n$ final-state particles. Each entry gives
  the \pyac{} time in black, in $\mu$s/sample, followed by
  $t_{\rm pyAC}/t_{\rm AC}$ in parentheses (orange above one, green below).
  \pyac{} reuses the all-helicity, all-flow process outputs and batched timing
  procedure of Table~\ref{tab:lc-matrix}; the faster recurrence
  JIT-O2 ($r$) or compiled JIT-O3 ($c$) result is selected.
  \ac{} uses the unchanged warmed single-helicity native interface of
  Table~\ref{tab:z-progression}. Numbers have three significant figures,
  with ratios computed before rounding.
  For non-$Z$ families, the seed alternates signs $(-,+,-,+,\ldots)$ in
  physical external-particle order, using transverse vector polarisations
  and zero scalar helicities. If needed, the fewest helicities are changed
  to obtain a nonzero matrix element; the selected configurations accompany
  the data. The $Z$-plus-jets row shares the single-helicity configurations
  and timings of Table~\ref{tab:z-progression}.
  In the four-lepton row, \pyac{} includes the correct identical-lepton
  interference, whereas \ac{} has a known interference-sign bug; the measured
  \ac{} timings and corresponding timing ratios are retained.
}
  \label{tab:lc-single-helicity}
  \small
  \renewcommand{\arraystretch}{1.12}
  \setlength{\tabcolsep}{3pt}
  \begin{tabularx}{\textwidth}{Y c c c c}
    \toprule
    Process family & $n=4$ & $n=5$ & $n=6$ & $n=7$ \\
    \midrule
    $d\bar d\to Z+(n-1)g$
      & \textcolor{black}{$0.222$}\,\textcolor{green!50!black}{$({\times} 0.510^{r})$} & \textcolor{black}{$0.360$}\,\textcolor{green!50!black}{$({\times} 0.456^{r})$} & \textcolor{black}{$0.602$}\,\textcolor{green!50!black}{$({\times} 0.454^{r})$} & \textcolor{black}{$0.993$}\,\textcolor{green!50!black}{$({\times} 0.477^{r})$} \\
    $u\bar d\to e^+\nu_e+(n-2)g$
      & \textcolor{black}{$0.162$}\,\textcolor{green!50!black}{$({\times} 0.634^{r})$} & \textcolor{black}{$0.315$}\,\textcolor{green!50!black}{$({\times} 0.661^{c})$} & \textcolor{black}{$0.513$}\,\textcolor{green!50!black}{$({\times} 0.623^{c})$} & \textcolor{black}{$0.648$}\,\textcolor{green!50!black}{$({\times} 0.482^{r})$} \\
    $d\bar d\to ZZ+(n-2)g$
      & \textcolor{black}{$0.182$}\,\textcolor{green!50!black}{$({\times} 0.387^{r})$} & \textcolor{black}{$0.294$}\,\textcolor{green!50!black}{$({\times} 0.371^{r})$} & \textcolor{black}{$0.490$}\,\textcolor{green!50!black}{$({\times} 0.395^{r})$} & \textcolor{black}{$0.772$}\,\textcolor{green!50!black}{$({\times} 0.401^{r})$} \\
    $gg\to t\bar t+(n-2)g$
      & \textcolor{black}{$0.474$}\,\textcolor{green!50!black}{$({\times} 0.812^{c})$} & \textcolor{black}{$0.831$}\,\textcolor{green!50!black}{$({\times} 0.797^{r})$} & \textcolor{black}{$1.30$}\,\textcolor{green!50!black}{$({\times} 0.742^{r})$} & \textcolor{black}{$1.95$}\,\textcolor{green!50!black}{$({\times} 0.763^{r})$} \\
    $gg\to gg+(n-2)g$
      & \textcolor{black}{$0.468$}\,\textcolor{green!50!black}{$({\times} 0.588^{r})$} & \textcolor{black}{$0.759$}\,\textcolor{green!50!black}{$({\times} 0.566^{r})$} & \textcolor{black}{$1.30$}\,\textcolor{green!50!black}{$({\times} 0.610^{r})$} & \textcolor{black}{$1.81$}\,\textcolor{green!50!black}{$({\times} 0.561^{r})$} \\
    $d\bar d\to e^+e^-ZH+(n-4)g$
      & \textcolor{black}{$0.347$}\,\textcolor{green!50!black}{$({\times} 0.583^{c})$} & \textcolor{black}{$0.460$}\,\textcolor{green!50!black}{$({\times} 0.546^{r})$} & \textcolor{black}{$0.559$}\,\textcolor{green!50!black}{$({\times} 0.478^{r})$} & \textcolor{black}{$0.728$}\,\textcolor{green!50!black}{$({\times} 0.440^{r})$} \\
    $d\bar d\to e^+e^-e^+e^-+(n-4)g$
      & \textcolor{black}{$0.386$}\,\textcolor{green!50!black}{$({\times} 0.473^{c})$} & \textcolor{black}{$0.606$}\,\textcolor{green!50!black}{$({\times} 0.547^{c})$} & \textcolor{black}{$0.754$}\,\textcolor{green!50!black}{$({\times} 0.512^{r})$} & \textcolor{black}{$0.926$}\,\textcolor{green!50!black}{$({\times} 0.461^{r})$} \\
    $d\bar d\to u\bar u s\bar s+(n-4)g$
      & \textcolor{black}{$0.225$}\,\textcolor{green!50!black}{$({\times} 0.682^{r})$} & \textcolor{black}{$0.383$}\,\textcolor{green!50!black}{$({\times} 0.792^{r})$} & \textcolor{black}{$0.592$}\,\textcolor{green!50!black}{$({\times} 0.843^{r})$} & \textcolor{black}{$1.00$}\,\textcolor{green!50!black}{$({\times} 0.973^{r})$} \\
    \bottomrule
  \end{tabularx}
  \par\vspace{0.6em}
  \begin{minipage}{\textwidth}
  \footnotesize
  \emph{Changes to the alternating seed.} Here $s_i=\operatorname{sign}(h_i)$
  is the sign of the physical helicity of leg $i$, with the incoming legs first.
  $e^+\nu_e+$jets ($n=4,5,6,7$): $s_{3}=+1,\,s_{4}=-1$.
  \end{minipage}
\end{table}
\setcounter{table}{\savedLCSingleHelicityTable}
\endgroup

\subsection{Full-colour comparisons}
\label{app:mg-comparison}

For full colour we compare two QCD families through five final-state
particles with \ac{} and a standalone \mgamc{}
calculation~\cite{Alwall:2014hca}.  We retain only QCD interactions,
setting \texttt{QED=0} in \pyac{} and \mgfive{}.  All calculations include
the complete colour and physical helicity sums, initial-state averages and
identical-particle symmetry factors.

Generation uses recurrence JIT-O2 and explicitly disables numerical
current-relation discovery. We compare trace and adaptive-adjoint FFT,
additionally testing direct contraction for the quark family, and report the
fastest validated mode. All tested modes agree with the independent reference
at three common regular phase-space points (seeds 101, 103 and 107), with
relative differences below $1.53\times10^{-13}$.

\begin{table}[H]
  \centering
  \caption{Full-colour, helicity-summed warmed evaluation times on one Apple M3 Pro
  CPU core, with $\alpha_s=0.118$ and $n$ final-state particles.
  \pyac{}~1.0.0 uses recurrence JIT-O2 in batches of 128;
  \mgamc{}~3.7.1 uses its generated single-point
  \texttt{SMATRIX} routine compiled with \texttt{-O3}, timed inside a native
  Fortran loop.  \ac{} uses its generated full-colour library, with amplitude
  evaluation and colour contraction both warmed before timing.
  \pyac{} and \mgfive{} use wall clocks; \ac{} uses single-core CPU time.
  A sample is one phase-space point; setup and warm-up are excluded.
    Superscripts A, T and D identify adaptive-adjoint FFT, trace FFT and direct colour contraction. Marked entries use the fastest validated mode tested for that process; A retains fundamental chains for quarks and uses DDM only where certified.
  Values have three significant
  figures; $R=t_{\rm pyAC}/t_{\rm AC}$ is calculated before rounding,
  with green below one and orange above one.
  }
  \label{tab:mg-comparison}
  \small
  \renewcommand{\arraystretch}{1.2}
  \setlength{\tabcolsep}{4pt}
  \begin{tabular*}{\textwidth}{@{\extracolsep{\fill}} l c
      S[table-format=5.3] S[table-format=5.3] S[table-format=5.3]
      S[table-format=1.3] @{}}
    \toprule
    Process family & $n$ & {\ac{}} & {\mgfive{}} & {\pyac{}} & {$R$} \\
      & & {[$\mu$s/sample]} & {[$\mu$s/sample]} & {[$\mu$s/sample]} & \\
    \midrule
    $d\bar d\to d\bar d+(n-2)g$ & 2 & 0.543 & 0.397 & \multicolumn{1}{c}{\tablenum[table-format=5.3]{0.209}\textsuperscript{A}} & \color{green!50!black} 0.386 \\
     & 3 & 3.02 & 3.75 & \multicolumn{1}{c}{\tablenum[table-format=5.3]{1.21}\textsuperscript{D}} & \color{green!50!black} 0.399 \\
     & 4 & 34.2 & 39.2 & \multicolumn{1}{c}{\tablenum[table-format=5.3]{8.36}\textsuperscript{T}} & \color{green!50!black} 0.245 \\
     & 5 & 517 & 670 & \multicolumn{1}{c}{\tablenum[table-format=5.3]{113}\textsuperscript{A}} & \color{green!50!black} 0.218 \\
    \addlinespace[0.65ex]
    $gg\to gg+(n-2)g$ & 2 & 4.08 & 3.00 & \multicolumn{1}{c}{\tablenum[table-format=5.3]{1.46}\textsuperscript{A}} & \color{green!50!black} 0.357 \\
     & 3 & 62.6 & 43.9 & \multicolumn{1}{c}{\tablenum[table-format=5.3]{7.58}\textsuperscript{A}} & \color{green!50!black} 0.121 \\
     & 4 & 1400 & 921 & \multicolumn{1}{c}{\tablenum[table-format=5.3]{92.9}\textsuperscript{A}} & \color{green!50!black} 0.0664 \\
     & 5 & 50600 & 46700 & \multicolumn{1}{c}{\tablenum[table-format=5.3]{1430}\textsuperscript{A}} & \color{green!50!black} 0.0283 \\
    \bottomrule
  \end{tabular*}
\end{table}

The comparison includes batching, colour contraction and the other differences
between the complete implementations. The measurement protocol and memory
limit are specified in appendix~\ref{app:benchmark-reproducibility};
the unrounded measurements, tested colour modes and phase-space inputs
accompany the paper.

\subsection{Reproducibility}
\label{app:benchmark-reproducibility}

The reference release for reproducing the calculations in this paper is
\pyac{}~1.0.0, which will be available on
\href{https://pypi.org/project/pyamplicol/}{PyPI} upon publication.
The \href{https://github.com/mg5amcnlo/pyamplicol}{source repository} and
\href{https://mg5amcnlo.github.io/pyamplicol/}{online manual} provide the
code and installation instructions. The data accompanying the tables record
the process definitions, model parameters, phase-space points, selected
helicities, unrounded measurements and detailed run metadata.
The built-in \texttt{profile} command
and campaign profiler described in appendix~\ref{app:profiling-campaigns}
provide the timing procedure; absolute times naturally depend on the machine.

Benchmarks run on one Apple M3 Pro CPU core. Unless noted otherwise, \pyac{}
times are warmed wall times per phase-space point, measured in native loops
with batches of 128, two warm-ups and at least five timed blocks. Profiling
targets are eight seconds for the $Z$ and HEFT families, ten seconds for the
full-colour QCD comparison and five seconds for the other selected-helicity
entries. Generation, loading and warm-up are excluded from the timings.
Local generation and evaluation are monitored by a 30\,GB process-tree
memory watchdog; generation has no time cutoff except for the compiled-mode
limits stated in the table captions.

The exact-colour benchmarks use recurrence JIT-O2, retain all helicities
and explicitly disable numerical current-relation discovery while preserving
structural sharing. We compare adjoint and trace FFT, and also direct
contraction for the quark and HEFT families, selecting the fastest complete
central timing estimate among the validated modes, not the fastest individual
block. The table superscripts identify that choice. All mode measurements
and their uncertainties remain in the data. The LC outputs likewise support
all helicities and flows, with any restriction imposed at evaluation time.

The largest HEFT process, $gg\to H+6g$, required generation on a larger-memory
cluster. Its portable JIT-O2 output was evaluated on the same Mac under the
30\,GB runtime guard, using one-point rather than 128-point batches.
Trace generation had already exceeded the local memory limit at $n=6$;
direct generation was not attempted at $n=6,7$ after the lower-multiplicity
comparisons. Missing modes are not assigned timings.

The independent \ac{} and \mgfive{} comparisons use matching physical
inputs and native scalar loops. For \ac{}, helicity sums use the generated
library, while selected-helicity comparisons supply helicities at runtime
through its generic single-flow interface. \ac{} reports single-core CPU time;
\mgfive{} reports wall time. Their Fortran code is compiled with
\gnufortran{}~14.2.0 and \texttt{-O3}, with \texttt{-ffast-math} additionally
used for \ac{}. We use \mgfive{}~3.7.1, except for the HEFT $n=7$
source generated with~3.7.0. Large \mgfive{} matrix routines are split into
separately compiled helpers without changing their algebra. For the expensive
HEFT $n=7$ comparison, one warm-up is followed by a single timed evaluation
of the full-colour helicity sum; unlike the other entries, it has no repeated
timing samples.

\Needspace{8\baselineskip}
\section{Correlated tree matrix elements}
\label{app:correlators}

Soft and collinear factorisation involves insertions in a reduced Born
amplitude~\cite{Catani:1996vz,Catani:1999ss}.  We describe the colour and spin
quantities supplied by \pyac{}, keeping the kinematic splitting functions
separate.  The number of colours is $N_c=3$; it is not the dimension
$N_{\rm basis}$ of the colour basis.  With external helicities denoted by $h$,
write
\begin{equation}
 |\mathcal M_h\rangle=\sum_{c=1}^{N_{\rm basis}} A_{hc}|C_c\rangle,
 \qquad
 B=\mathcal N\sum_{h,c,d}A_{hc}^*\,
       \langle C_c|C_d\rangle\,A_{hd}.
 \label{eq:correlated-basis}
\end{equation}
Here $|\mathcal M_h\rangle$ is the tree amplitude for helicity assignment $h$, viewed as a vector in colour space, $|C_c\rangle$ are the colour basis vectors, labelled by $c$ and $d$, $\langle C_c|C_d\rangle$ is their overlap matrix (the matrix $G$ of section~\ref{sec:colour-sums}, whose basis index is written $c$ here) and $B$ is the Born matrix element squared, summed over colours and helicities.  The partial amplitudes $A_{hc}$ retain their relative phases.  The common
normalisation $\mathcal N$ includes the usual incoming-state averages and
any final-state symmetry factor.  It is used consistently for the Born
quantity and every insertion; it contains no unresolved phase space or
additional splitting coupling.

\subsection{Colour charges and ordered connections}

We use Hermitian fundamental generators $t^a$, with $a,b,c$ adjoint indices and $f^{abc}$ the real, totally antisymmetric structure constants, satisfying
$\operatorname{tr}(t^a t^b)=T_R\delta^{ab}$, $T_R=1/2$, and
$[t^a,t^b]=if^{abc}t^c$.  In output--input index order, with $i,j$ fundamental and $b,c$ adjoint colour indices, the charge acting
on a quark, antiquark or gluon is, respectively,
\begin{equation}
 (\mathbf T_q^a)_{ij}=t^a_{ij},\qquad
 (\mathbf T_{\bar q}^a)_{ij}=-t^a_{ji},\qquad
 (\mathbf T_g^a)_{bc}=-if^{abc}.
 \label{eq:correlated-charges}
\end{equation}
Colour charges are all outgoing: an incoming quark carries the antiquark
charge and conversely.  This crossing is performed by \pyac{}; public leg
labels and the physical input momenta retain their original ordering.
The corresponding Casimirs are $C_F=(N_c^2-1)/(2N_c)$ and $C_A=N_c$.
The conversion from any internally rescaled generators to the normalisation
in eq.~\eqref{eq:correlated-charges} is included in the colour matrices.

Let $S_\alpha^{(m)}$, labelled by $\alpha$, be a sequence of $m$ colour
emissions and splittings.  Each operation acts on a current parton label:
gluon emission applies its colour charge and introduces an adjoint index;
a gluon-to-quark-pair splitting replaces an adjoint index by a fundamental
and an antifundamental index through $t^a_{q\bar q}$, with $a$ the adjoint index of the split gluon and $q,\bar q$ the daughter colour indices.  An emitted parton
may participate in a later operation.  For a chronological sequence
$O_1,\ldots,O_m$ of such operations, the ket operator is the product $S_\alpha^{(m)}=O_m\cdots O_1$, in which $O_1$ acts first and later operations stand to the left.
Thus successive emissions from the same line retain their order, while
charges on distinct lines commute.

To generate these connections, start with the identity and recursively
attach one further emission to every active coloured leg.  Every emitted
gluon can in turn radiate or split into a quark--antiquark pair, whose
daughters can radiate at subsequent steps.  Denote the active coloured legs
after a history $S$ by $L(S)$, and its still-unsplit auxiliary gluons by $G(S)\subseteq L(S)$.  With
$E_i$ denoting gluon emission from leg $i$ and $Q_g$ quark-pair splitting
of gluon $g$, each written to the left of the history $S$ it extends, the sets $\mathcal S_m$ of all histories with $m$ operations are generated from the identity $1$ by the recursion
\begin{equation}
 \mathcal S_0=\{1\},\qquad
 \mathcal S_{m+1}
 =\bigcup_{S\in\mathcal S_m}
   \left(\{E_i S:i\in L(S)\}\cup\{Q_g S:g\in G(S)\}\right).
 \label{eq:colour-connection-catalogue}
\end{equation}
All assignments of the $m$ final unresolved labels are included.
In this soft construction a hard Born gluon is not split into a quark pair;
such a splitting can also be declared explicitly when needed.

The histories form a spanning set, not a minimal basis.  They include
independent hard-line emissions, ordered emissions on a shared line and
all gluon or quark cascades rooted in an emitted gluon.  The colour factors
of a four-gluon vertex are products of two structure constants and hence
are already spanned by successive cubic colour operations.  This gives a
recursive construction of the colour tensors of leading tree-level QCD soft
currents, without first reducing them to products of dipoles.  The chronology
orders colour operations, not soft energies: simultaneous-soft kinematics
remain in the coefficients multiplying these tensors~\cite{Catani:2022soft}.
For example, three soft gluons already generate colour quadrupoles in
addition to dipole correlations~\cite{Catani:2019triple}.

Pairs of same-order connections $\alpha,\beta$ with the same labelled final colour space, that is the same open colour indices of the final unresolved partons, define the inserted colour matrix $K^{\alpha\beta}_{cd}$, the overlap of the basis vectors acted on by the two connections, and the corresponding colour-correlated Born quantity $B_{\alpha\beta}$,
\begin{equation}
 K^{\alpha\beta}_{cd}
   =\langle S_\alpha C_c|S_\beta C_d\rangle,
 \qquad
 B_{\alpha\beta}=\mathcal N\sum_{h,c,d}
                  A_{hc}^*K^{\alpha\beta}_{cd}A_{hd}.
 \label{eq:ordered-colour-connections}
\end{equation}
Here the common order $m$ is left implicit.  The bra connection is adjointed.  Therefore
$K^{\beta\alpha}=(K^{\alpha\beta})^\dagger$ and
$B_{\beta\alpha}=B_{\alpha\beta}^*$; an individual ordered contraction can
be complex.  At full colour a self-contraction is a norm.  Empty connections
recover eq.~\eqref{eq:correlated-basis}.  At NLO, one emission from hard leg $i$ on the bra side and from hard leg $j$ on the ket side gives the colour dipole $B_{ij}$, where $\mathbf T_i^a$ is the charge of eq.~\eqref{eq:correlated-charges} appropriate to leg $i$ and the dot contracts the adjoint index,
\begin{equation}
 B_{ij}=\mathcal N\sum_h
       \langle\mathcal M_h|\mathbf T_i\cdot\mathbf T_j|\mathcal M_h\rangle,
 \qquad \mathbf T_i\cdot\mathbf T_j=\sum_a\mathbf T_i^a\mathbf T_j^a.
 \label{eq:colour-connected-born}
\end{equation}
A dipole $B_{ij}$ need not be positive.  Diagonal entries are retained,
although their soft eikonal coefficient vanishes for a massless emitter.

If a physical current $J^{(m)}$ for $m$ unresolved partons is expanded in the connections, with kinematic coefficients $j_\alpha$, as $J^{(m)}=\sum_\alpha j_\alpha
S_\alpha^{(m)}$, its squared contribution is
$\sum_{\alpha\beta}j_\alpha^*j_\beta B_{\alpha\beta}$, with any unresolved
spin sums understood.  Thus all compatible ordered pairs expose the
interferences needed for any projection onto another colour basis.
Gluon-only and quark-pair final spaces form separate blocks.  Their label
assignments are matched before pairing, including histories in which
the same final partons were created in a different order.

For a concrete N$^3$LO interference, consider four distinct hard lines
$i,j,k,l$, with $i,k$ quarks and $j,l$ antiquarks, and three unresolved
gluons of colours $a_1,a_2,a_3$.  In connection $\alpha$, line $i$ emits a
gluon of colour $b$, which radiates gluon $a_1$ and changes colour to $c$,
then radiates gluon $a_2$ and ends with colour $a_3$.  In connection $\beta$,
the same three final gluons are emitted independently from lines $j,k,l$.
Suppressing identity operators on untouched hard lines, and writing the open adjoint indices of the three final gluons as superscripts in place of the order label $(m)$, these histories give
\begin{equation}
 \begin{aligned}
 S_\alpha^{a_1a_2a_3}
   &=\sum_{b,c}(-if^{a_2a_3c})(-if^{a_1cb})\,\mathbf T_i^b,
 \\
 S_\beta^{a_1a_2a_3}
   &=\mathbf T_j^{a_1}\mathbf T_k^{a_2}\mathbf T_l^{a_3}.
 \end{aligned}
 \label{eq:n3lo-interference-histories}
\end{equation}
The $\mathbf T$ operators here are fundamental generators $t^a$ on the
quarks and $-(t^a)^T$ on the antiquarks, whereas the two $f$ tensors describe
the adjoint cascade.  Contracting the same three unresolved colours between
the two sides gives
\begin{equation}
 B_{\alpha\beta}
 =-\mathcal N\!\sum_{h,a_1,a_2,a_3,b,c}
 f^{a_2a_3c}f^{a_1cb}
 \langle\mathcal M_h|
 \mathbf T_i^b\mathbf T_j^{a_1}\mathbf T_k^{a_2}\mathbf T_l^{a_3}
 |\mathcal M_h\rangle.
 \label{eq:n3lo-interference-quadrupole}
\end{equation}
This is a four-line colour correlation, or quadrupole, obtained from two
different three-emission histories rather than the square of either one.
The minus sign follows from the two adjoint charges in
eq.~\eqref{eq:correlated-charges}.  In the physical squared soft current,
this entry and its reverse supply
$2\operatorname{Re}(j_\alpha^*j_\beta B_{\alpha\beta})$.
Thus the catalogue exposes the interference itself, leaving its kinematic
coefficient to the subtraction prescription.

There is no fixed limit on the number of operations.  The same construction
applies at NLO, NNLO, N$^3$LO and higher orders; its cost grows with the number
of histories and their pairings.  It supplies the tree-level colour
structures, not loop amplitudes or kinematic subtraction terms.
The quark-pair operation describes one labelled flavour,
with no implicit sum over flavours.  At full colour its normalisation follows from the generator trace: with $(i\to g\to q\bar q)$ denoting the history in which leg $i$ emits a gluon that then splits into the labelled quark pair, the same history on both sides gives
\begin{equation}
 \sum_{q,\bar q}(t^a_{q\bar q})^*t^b_{q\bar q}
      =T_R\delta^{ab},\qquad
 B_{(i\to g\to q\bar q),(i\to g\to q\bar q)}=T_R B_{ii}.
 \label{eq:correlated-pair-trace}
\end{equation}

The automatic generation option applies
eq.~\eqref{eq:colour-connection-catalogue} through the requested order and
registers every compatible pair.  It identifies reorderings of independent
operations, retaining chronological execution and the order along a shared
line.  Relations following from colour conservation or opposite signs are
not used to reduce the list.  Each connection receives a deterministic index
within its order; a correlator ID records the order and its bra and ket
indices.  The full ordered histories remain available for inspection at
runtime, as illustrated in appendix~\ref{app:minimal-manual}.  No symmetry
multiplicity is folded into an individual $B_{\alpha\beta}$: weights belong
to the chosen physical current or colour-conservation sum.

Ordered connections also retain products sharing a hard colour line.  For
example, the double-soft current contains the anticommutator of two dipole operators,
\begin{equation}
 \mathcal N\sum_h\langle\mathcal M_h|
 \{\mathbf T_i\cdot\mathbf T_j,\mathbf T_k\cdot\mathbf T_l\}
 |\mathcal M_h\rangle,\qquad \{X,Y\}=XY+YX.
 \label{eq:correlated-anticommutator}
\end{equation}
The two orders are evaluated separately and added, with no factor of $1/2$.
They must not be identified when a colour line is shared.  We construct
such products as colour tensors, not by multiplying overlap matrices in
the generally nonorthogonal Born basis.

\subsection{Colour accuracy}

Connection order and colour accuracy are independent choices.  LC, NLC
and full-colour correlated quantities use the complete generated amplitude
basis, with the requested accuracy applied to the inserted matrix.
For a basis containing $n_g$ adjoints and $n_{q\bar q}$ fundamental pairs,
a connected family with $r$ operations including $s$ quark-pair splittings
on each side has, as reference power of $N_c$,
\begin{equation}
 P=n_g+n_{q\bar q}+r-s.
 \label{eq:correlated-colour-power}
\end{equation}
LC retains powers of $N_c$ at least $P$.  For a purely adjoint final colour
space, NLC retains powers at least $P-2$.  With fundamental lines it retains
the exact coefficients of entries reaching that order, following the
ordinary open-line convention of section~\ref{sec:colour-sums}.  Intervening odd powers are included.
The empty connection uses the ordinary Born prescription.  Amplitude
coefficients and incoming averages remain at physical $N_c=3$: this is a
colour-matrix approximation, not a new expansion of the entire amplitude.
The reference power is common to related insertions and is not reset for
an individually suppressed dipole.  Full colour retains all terms.

\subsection{Spin insertions and validation}

For one vector leg, replace its polarisation vector by an arbitrary source vector $v^\mu$ and write the resulting linear dependence as
$\mathcal M_h(v)=v^\mu\mathcal M_{h,\mu}$, where $\mathcal M_{h,\mu}$ is the amplitude with the source vector stripped and $h$ now labels the
unreplaced helicities.  For any colour insertion $K$, with $A_{hc,\mu}$ the correspondingly stripped partial amplitudes of eq.~\eqref{eq:correlated-basis}, define the spin-correlated tensor $B_{\mu\nu}[K]$ and its contraction $B(v;K)$ with the source vector and its conjugate,
\begin{equation}
 B_{\mu\nu}[K]=\mathcal N\sum_{h,c,d}
          A_{hc,\mu}^*K_{cd}A_{hd,\nu},\qquad
 B(v;K)=v^{\mu*}v^\nu B_{\mu\nu}[K].
 \label{eq:spin-correlated-born}
\end{equation}
The supplied contravariant vector replaces the external source with metric
$(+,-,-,-)$, without normalisation, conjugation or transverse projection.
Several vectors act jointly on the same amplitude.  These four-dimensional
rank-one contractions suffice for the splitting term in
appendix~\ref{sec:infrared-stability}; no additional spin average is applied.
Full density tensors and independent bra and ket vectors are not returned.

We test ordinary Born recovery, the Casimir and colour-conservation
identities
\begin{equation}
 B_{ii}=C_iB,\qquad
 \sum_{j\,\mathrm{coloured}}B_{ij}=0,
 \label{eq:correlated-coherence}
\end{equation}
including $j=i$; here $C_i$ is the Casimir of leg $i$, $C_F$ for a quark or antiquark and $C_A$ for a gluon.  Colour conservation applies to all active partons,
including emitted ones but not parents replaced by splitting.  Tests cover
three operations and additional higher-order Casimir and two-quark-pair
identities.  We compare the catalogue with explicit low-order enumeration and
histories reaching the same final space in different orders.  LC/NLC identities
hold through the retained order; extra terms kept by the exact-coefficient
NLC prescription need not cancel beyond it.

For spin, two orthonormal transverse vectors satisfy
$B(e_1;K)+B(e_2;K)=B[K]$ for a massless vector leg, where $B[K]$ is the
physical-helicity sum with insertion $K$.  Replacing a polarisation
by its on-shell momentum gives the Ward zero, with the remaining gauge legs
physical; several simultaneous momentum replacements are tested as well.
We also test incoming-leg crossing, complex vector rescaling, noncommuting
colour products and individual/grouped agreement across phase-space points.

\clearpage
\section{Numerical stability of local subtraction}
\label{sec:infrared-stability}

Local subtraction requires accurate differences between real-emission matrix
elements and their counterterms. We examine $gg\to gggg$ in two separate
limits: a final-state collinear pair and a wide-angle soft gluon. Write $R$
for the real squared matrix element, $B$ for the mapped $gg\to ggg$ Born,
$C$ for the corresponding local counterterm, and $F=C-R$ for the tested
difference. All quantities include full colour,
summed helicities of every leg whose polarisation is not replaced by an explicit vector in the spin-correlated Born and the same initial-state averages, but no
final-state symmetry factors. The Born insertions are those of
appendix~\ref{app:correlators}.

\subsection{Local collinear and soft counterterms}

For gluons $i,j$ becoming collinear, the four-dimensional splitting
tensor~\cite{Catani:1996vz,Catani:1999ss} gives the collinear counterterm
\begin{equation}
 C_{\rm coll}=\frac{8\pi\alpha_s}{s_{ij}}\,2C_A
 \left[\left(\frac{z}{1-z}+\frac{1-z}{z}\right)B
       +2z(1-z)B(e)\right],\qquad
 e^\mu=\frac{k_\perp^\mu}{\sqrt{-k_\perp^2}}.
 \label{eq:collinear-counterterm}
\end{equation}
Here $C_A=3$ is the adjoint Casimir, $s_{ij}=(p_i+p_j)^2$ is the invariant mass squared of the pair with momenta $p_i,p_j$, $z$ is the momentum fraction of gluon $i$ (eq.~\eqref{eq:infrared-map}), $k_\perp$ is the transverse momentum of the splitting relative to the parent direction, also defined there, and $B(e)$
replaces the parent gluon's polarisation by the real transverse unit vector
$e$. Its coefficient is positive for the $(+,-,-,-)$ metric. Azimuthal
averaging gives $\langle B(e)\rangle=B/2$ and the spin-averaged splitting
function. We instead retain the actual azimuth: the counterterm is local
without pairing rotated events.

We keep a generic Born point at $Q=\sqrt{s}=1\,\mathrm{TeV}$ fixed, where $s$ is the partonic centre-of-mass energy squared and $Q$ sets the hard scale.
For null mapped emitter and spectator momenta $P,K$, the inverse final-final
Catani--Seymour map is
\begin{align}
 p_i&=zP+y(1-z)K+k_\perp,&
 p_j&=(1-z)P+yzK-k_\perp,& p_k&=(1-y)K,\nonumber\\
 y&=\frac{k_T^2}{2z(1-z)P\cdot K},&
 k_T&=\sqrt{-k_\perp^2},& s_{ij}&=\frac{k_T^2}{z(1-z)}.
 \label{eq:infrared-map}
\end{align}
Here $p_k$ is the real-emission momentum of the spectator gluon $k$, whose mapped momentum is $K$. The vector $k_\perp$ is orthogonal to $P,K$; other momenta are unchanged.
This conserves momentum and keeps every leg on shell: $k_T$ is the magnitude of the transverse momentum, and the dimensionless Catani--Seymour variable $y$ takes the value that makes $p_i$ and $p_j$ massless. The collinear
trajectory is parametrised by a dimensionless scaling variable $\delta$, with $\delta\to0$ the singular limit: it uses $z=3/5$, $k_T=\delta Q$ and the generic direction
$e=(3e_1+4e_2)/5$, with $e_1,e_2$ an orthonormal basis of the plane transverse to $P$ and $K$. Neither daughter
becomes soft and no other collinear limit is approached.

For a soft gluon of momentum $q$, the counterterm is
\begin{equation}
 C_{\rm soft}=-8\pi\alpha_s
 \sum_{a<b}\frac{P_a\cdot P_b}{(P_a\cdot q)(P_b\cdot q)}B_{ab}.
 \label{eq:soft-counterterm}
\end{equation}
The sum counts each unordered pair of coloured Born legs once, including
incoming legs. Here $B_{ab}$ is the colour-correlated Born of eq.~\eqref{eq:colour-connected-born}, with the colour-charge product $\mathbf T_a\cdot\mathbf T_b$ of legs $a$ and $b$ inserted. The ten dipoles use the crossed colour charges already in
$B_{ab}$, without further signs. Eikonal factors use mapped Born momenta
$P_a$; diagonal terms vanish for these massless partons~\cite{Catani:1999ss}.

Choose a fixed null vector $r$ at wide angle to all hard partons and set
$q=\delta r$, with the same scaling variable $\delta$. Decompose $r$ along the mapped emitter and spectator momenta as $r=\alpha P+\beta K+r_\perp$, where $r_\perp$ is the component orthogonal to both $P$ and $K$ and the coefficients are
$\alpha=r\cdot K/(P\cdot K)$ and $\beta=r\cdot P/(P\cdot K)$.
Equation~\eqref{eq:infrared-map}, with $z=1-\delta\alpha$,
$y=\delta\beta/z$ and $k_\perp=-\delta r_\perp$, gives $p_j=q$ exactly; this $y$ equals the on-shell value in eq.~\eqref{eq:infrared-map} because $r^2=0$ gives $-r_\perp^2=2\alpha\beta\,P\cdot K$
and retains the same mapped Born. We use $\alpha_s=1/8$, exactly
representable in the tested arithmetics. These are two separate local-limit
tests, not counterterms combined into a global subtraction prescription.

\subsection{Precision comparison and reproducibility}

\begin{figure}[H]
 \centering
 \includegraphics[width=\textwidth]{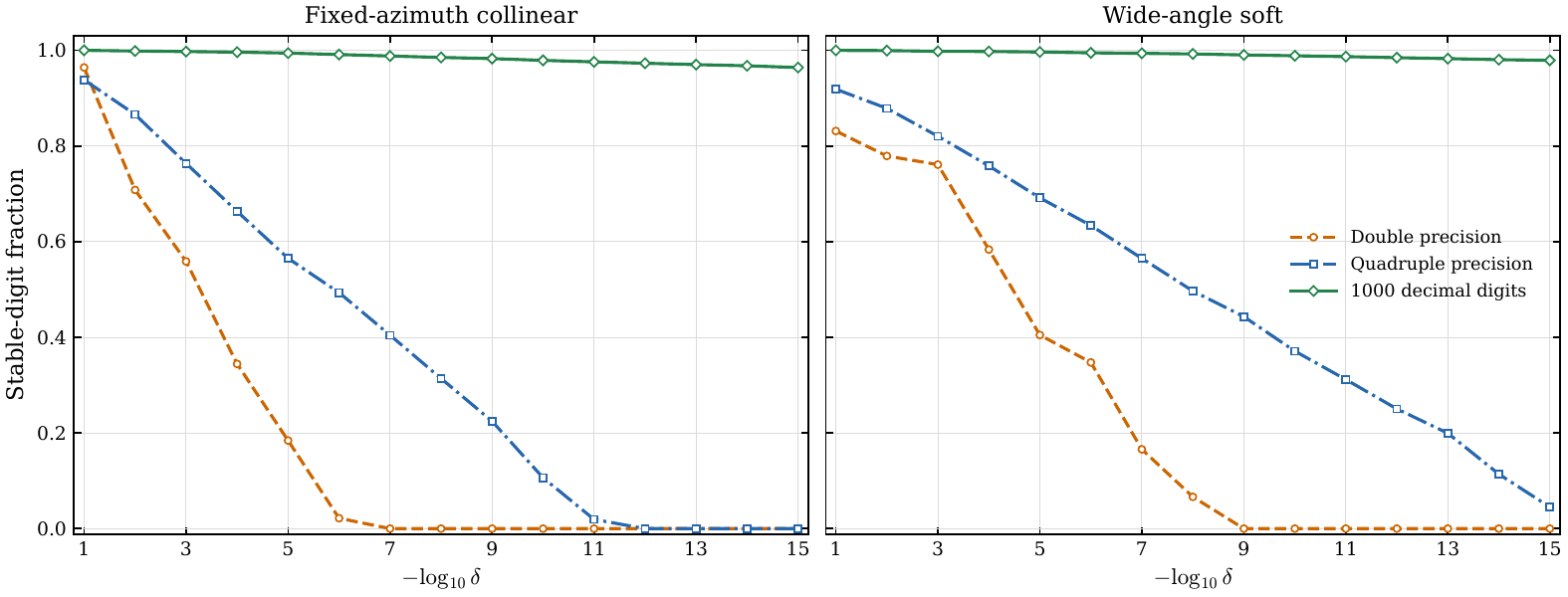}
 \caption{Retained decimal-digit fraction in the full-colour $gg\to gggg$
 difference $F=C-R$, with the spin-correlated collinear counterterm (left)
 and colour-connected soft counterterm (right) of
 eqs.~\eqref{eq:collinear-counterterm} and \eqref{eq:soft-counterterm}.
 The stability measure is defined in eq.~\eqref{eq:infrared-stability}.
 Each marker is an evaluated integer $-\log_{10}\delta=1,\ldots,15$;
 lines guide the eye. Zero means no stable digit.}
 \label{fig:collinear-stability}
\end{figure}

Inputs are constructed with guard digits, then rounded into the arithmetic
used throughout the matrix elements, counterterms and subtraction.
A 1200-digit reference $F_\star$ uses the unrounded trajectory. We measure
the fraction of retained decimal digits,
\begin{equation}
 \epsilon_F=\frac{|F-F_\star|}{|F_\star|},\qquad
 f_{\rm stable}=\min\!\left(1,
 \frac{\max(0,-\log_{10}\epsilon_F)}{d}\right).
 \label{eq:infrared-stability}
\end{equation}
Here $\epsilon_F$ is the relative error of $F$ with respect to the reference, $f_{\rm stable}$ the fraction of the $d$ nominal decimal digits of the arithmetic that survive, and $d=53\log_{10}2$ (the 53-bit binary64 significand), 31 and 1000 for the three curves. The double-double curve (labelled
``quadruple precision'' in fig.~\ref{fig:collinear-stability}) uses \symbolica{}'s DoubleFloat double-double arithmetic
with 31-digit Python transfers,
not the guarded multiprecision mode of section~\ref{sec:symbolica}.
The binary64 calculation also evaluates the correlated Born in double
precision. Checks at 1400 digits confirm negligible reference errors at
$\delta=10^{-15}$ and near the points where half the decimal digits are
lost, quoted below.

We check the map, polarisation completeness and colour coherence, and verify
$C/R\to1$ in both limits. Here $R,C=\mathcal O(\delta^{-2})$ but
$F=\mathcal O(\delta^{-1})$; this subleading singularity is integrable with
the unresolved phase-space measure.
Figure~\ref{fig:collinear-stability} shows the retained decimal-digit
fraction along both trajectories.

For this Born point, spin averaging leaves an asymptotic relative mismatch
of $4.85\times10^{-4}$; spin correlations restore local factorisation.
At $\delta=10^{-15}$, the 1000-digit results retain 964 and 979 decimal
digits in the collinear and soft limits, respectively. For this 1000-digit
calculation, half the decimal digits are lost between $\delta=10^{-169}$
and $10^{-170}$ (collinear), and between
$10^{-254}$ and $10^{-255}$ (soft), for these two trajectories.

The \href{https://github.com/mg5amcnlo/pyamplicol/blob/main/examples/NLO_limits/nlo_limit_stability_tests.py}{standalone NLO stability script}
reproduces generation, both batched scans and this figure, requesting all
ten dipoles and the spin projection together. Real current reuse is enabled;
correlated Born reductions follow section~\ref{sec:correlated-design}.
The reference release is specified in appendix~\ref{app:benchmark-reproducibility}.

Stable formulations of the kinematics and counterterms complement increased
precision~\cite{Bothmann:2024hgq}.

\clearpage
\Needspace{8\baselineskip}
\section{Minimal manual}
\label{app:minimal-manual}

This appendix gives a short route from installation to a numerical
matrix element.  Python~3.11 or later is required.  The complete manual,
including model support, configuration files, native interfaces and profiling,
is maintained in the
\href{https://mg5amcnlo.github.io/pyamplicol/}{\pyac{} documentation}.

\subsection{Installation and example output}

The examples use \pyac{}~1.0.0, the reference release for this paper.
Once it is published on PyPI, install it in a virtual environment and copy
the distributed examples as follows. Until then, use the source-installation
instructions in the \href{https://mg5amcnlo.github.io/pyamplicol/user/installation.html\#contributor-installation}{online manual}.

\begin{codecell}
\begin{lstlisting}[style=shellcode]
python3 -m venv .venv
. .venv/bin/activate
python -m pip install --upgrade pip
python -m pip install pyamplicol==1.0.0
pyamplicol examples copy ./pyamplicol-examples --force
cd pyamplicol-examples
\end{lstlisting}
\end{codecell}

\Needspace{12\baselineskip}
The supplied $pp\to Zjj$ example can then be generated, inspected and
evaluated as follows:

\begin{codecell}
\begin{lstlisting}[style=shellcode]
pyamplicol generate_pp_zjj_from_ufo_sm.toml
pyamplicol inspect artifacts/pp_zjj \
  --process 'd d~ > g z g'
pyamplicol evaluate artifacts/pp_zjj \
  --process 'd d~ > g z g' \
  --momenta data/pp_zjj_momenta.json \
  --model-parameters data/model_parameters.json
\end{lstlisting}
\end{codecell}

The supplied example uses leading colour; the last command sums its
colour-flow contributions and helicities.  Adding
\texttt{--resolved} retains the separate components; the equivalent prepared
commands are \texttt{pyamplicol evaluate\_total.toml} and
\texttt{pyamplicol evaluate\_resolved.toml}.  Momenta are supplied in process
order, with one row $(E,p_x,p_y,p_z)$ per external particle and an optional
outer axis for several phase-space points.  Incoming momenta have their
physical positive energies, not the all-outgoing signs used in the recursion.

\Needspace{8\baselineskip}
For NLC and full-colour recurrence or on-the-fly output, generation uses
the symmetric-group FFT with an adaptive colour basis by default when
supported, and otherwise falls back to direct contraction. The example below
requests the adaptive FFT explicitly:

\begin{codecell}
\begin{lstlisting}[style=shellcode]
pyamplicol generate 'g g > g g g' processes/ggg_fft \
  --model built-in-sm --color-accuracy full \
  --execution-mode recurrence \
  --fft adjoint
\end{lstlisting}
\end{codecell}

This uses the DDM basis described in section~\ref{sec:colour-sums} for the
gluon process. For other processes the adaptive choice retains the suitable
fundamental-chain or trace representation. Use \texttt{--fft trace} to
request FFT in the original basis, or \texttt{--color-contraction direct}
to disable it. LC, compiled, eager and correlated outputs use direct
contraction.

\Needspace{17\baselineskip}
\subsection{Python evaluation}

The same completed process output can be loaded directly from Python:

\begin{codecell}
\begin{lstlisting}[style=pythoncode]
import json
from pathlib import Path
from pyamplicol import Runtime

points = json.loads(Path("data/pp_zjj_momenta.json").read_text())
matrix_element = Runtime.load(
    "artifacts/pp_zjj", process="d d~ > g z g",
    model_parameters={"aS": 0.117, "MZ": 91.1876},
)
values = matrix_element.evaluate(points)
components = matrix_element.evaluate_resolved(points)
\end{lstlisting}
\end{codecell}

Here \texttt{evaluate} returns the requested physical sum, while
\texttt{evaluate\_resolved} retains its helicity components and, at LC, its
individual colour-flow contributions.  At NLC and full colour the colour axis
contains the contracted result.
Generated outputs also provide examples calling the same native
double-precision evaluator from C, C++, Fortran and Rust. The installed package
supplies the headers and language wrappers; further examples are given in the
\href{https://mg5amcnlo.github.io/pyamplicol/}{online manual}.

\Needspace{12\baselineskip}
\subsection{Correlated matrix elements}

We explicitly register the N$^3$LO interference of
eqs.~\eqref{eq:n3lo-interference-histories}--\eqref{eq:n3lo-interference-quadrupole}
for $d\bar d\to d\bar d g$.  Its four hard lines are
$(i,j,k,l)=(2,1,3,4)$: crossing the incoming $d,\bar d$ gives an outgoing
antiquark and quark, respectively.  The Born gluon on leg~5 is untouched by
the colour histories; we also allow its polarisation to be replaced.
Negative labels $-1,-2,-3$ identify the three auxiliary gluons of colours
$a_1,a_2,a_3$, not extra input momenta:

\begin{codecell}
\begin{lstlisting}[style=pythoncode]
from pyamplicol import (
    ColorCorrelator, CorrelatorConfig, CorrelatedRequest, EmitGluon,
    Generator, ModelSource, Runtime,
)
from pyamplicol.config import ColorConfig, RunConfig

alpha = (EmitGluon(2, -3), EmitGluon(-3, -1), EmitGluon(-3, -2))
beta = (EmitGluon(1, -1), EmitGluon(3, -2), EmitGluon(4, -3))
declarations = CorrelatorConfig(
    color_correlations=(
        ColorCorrelator("cascade-interference", bra=alpha, ket=beta),
    ),
    spin_correlations=((5,),),
)
# Alternatively, register every colour correlation through N3LO:
# declarations = CorrelatorConfig.all_color(
#     through_order=3, spin_correlations=((5,),),
# )
\end{lstlisting}
\end{codecell}

\Needspace{10\baselineskip}
Pass the declaration to the generator and load the resulting process output:

\begin{codecell}
\begin{lstlisting}[style=pythoncode]
config = RunConfig(action="generate", color=ColorConfig(accuracy="full"))
output = Generator(config).generate(
    "d d~ > d d~ g", "processes/correlated",
    model=ModelSource.built_in_sm(), correlators=declarations,
)
runtime = Runtime.load(output.output)
\end{lstlisting}
\end{codecell}

\Needspace{25\baselineskip}
Enumerate the available IDs and their ordered histories, then evaluate
them together over a batch \texttt{points} of momenta in this process order:

\begin{codecell}
\begin{lstlisting}[style=pythoncode]
incoming = ((400, 0, 0, 400), (400, 0, 0, -400))
points = (
    (*incoming, (300, 300, 0, 0), (250, -150, 200, 0),
     (250, -150, -200, 0)),
    (*incoming, (300, 0, 300, 0), (250, 200, -150, 0),
     (250, -200, -150, 0)),
)
catalogue = runtime.available_color_correlations()
for entry in catalogue:
    print(entry.id, entry.order, entry.bra, entry.ket)
requests = {
    entry.id: CorrelatedRequest(entry.id, spin_vectors={})
    for entry in catalogue
}
requests["spin"] = CorrelatedRequest(
    "cascade-interference", {5: (0, 0, 0, 1)},
)
values = runtime.evaluate_correlated_many(points, requests, precision=40)
entry = values["cascade-interference"][0]
print(entry.real, entry.imag)
\end{lstlisting}
\end{codecell}

Here \texttt{cascade-interference} is the ID chosen at generation;
\texttt{born} is always available.  The first result index labels the requested
contraction; the second follows the phase-space input order.
The commented alternative registers the complete set through order $k$:
$k=1$ for NLO, $k=2$ through NNLO, $k=3$ through N$^3$LO, and similarly
at higher orders.  Its generated IDs \texttt{N\{r\}LO/c\{i\}/c\{j\}}
give the order and connection indices; the printed histories give their
meaning.  With that alternative, use one of those IDs in the spin request
and final lookup.
The complete set grows rapidly, whereas the explicit declaration prepares
only the chosen interference and the Born overlap.
The request \texttt{spin} combines the same colour interference with the
supplied vector on leg~5.  Spin vectors can instead have one value per point.
The setter \texttt{set\_spin\_correlation\_vectors} supplies a default for
subsequent correlated calls.  The metric, normalisation, LC/NLC choices and ordered
connections are defined in appendix~\ref{app:correlators};
the \href{https://mg5amcnlo.github.io/pyamplicol/correlators.html}{online guide}
gives further examples.

\Needspace{15\baselineskip}
For high-precision spin vectors, construct \texttt{Decimal} components from
strings.  This avoids a preliminary rounding to binary64.  The same requests
can be evaluated in arbitrary precision or explicitly in double-double:

\begin{codecell}
\begin{lstlisting}[style=pythoncode]
from decimal import Decimal as D

vector = (D("0"), D("0"), D("0"), D("1.0000000000000000000000001"))
requests["spin"] = CorrelatedRequest("cascade-interference", {5: vector})
arb = runtime.evaluate_correlated_many(points, requests, precision=80)
dd = runtime.evaluate_correlated_many(
    points, requests, precision=31, arithmetic="double-double",
)
\end{lstlisting}
\end{codecell}

A complex component can be supplied as a pair of real and imaginary
\texttt{Decimal} values.  The native C, C++, Fortran and Rust interfaces
provide the same colour IDs, spin defaults and grouped evaluations in
binary64.  Their results retain both the request and phase-space indices;
complete examples are supplied in \texttt{examples/native/correlated.*}.
The \href{https://mg5amcnlo.github.io/pyamplicol/user/native-apis.html}{native API guide}
gives the language-specific calling conventions.

\Needspace{10\baselineskip}
The standalone script used in appendix~\ref{sec:infrared-stability} generates
the real and correlated Born processes, constructs the local counterterms,
evaluates the limits in batches and renders the figure.  From the repository
checkout, run:

\begin{codecell}
\begin{lstlisting}[style=shellcode]
python examples/NLO_limits/nlo_limit_stability_tests.py --smoke
python examples/NLO_limits/nlo_limit_stability_tests.py
python examples/NLO_limits/nlo_limit_stability_tests.py --render
\end{lstlisting}
\end{codecell}

Completed measurements are retained; the last command only redraws the
available results.  The example's README describes precision settings,
memory limits and smaller scan selections.

\Needspace{12\baselineskip}
\subsection{Profiling one process}

The same process output can be timed in batches, excluding generation:

\begin{codecell}
\begin{lstlisting}[style=shellcode]
pyamplicol profile artifacts/pp_zjj \
  --process 'd d~ > g z g' --batch-size 128 \
  --warmup-runs 2 --minimum-samples 5 --target-runtime 5
\end{lstlisting}
\end{codecell}

This requests two warm-up calls and at least five timed samples, targeting
five seconds of evaluation.  The report gives per-point timings and identifies
the colour and helicity workload chosen for this process output.
\texttt{--helicity} and \texttt{--color-flow} select particular components;
\texttt{--json} returns machine-readable results.

\Needspace{20\baselineskip}
\subsection{Profiling campaigns}
\label{app:profiling-campaigns}

To run the full built-in measurement campaign, use the profiler shipped with
\pyac{}. Supply local \ac{} and \mgfive{} directories as below.
The \ac{} checkout needs the profiling routines provided by its
\texttt{amplicol\_with\_patches} branch.  The comparisons require C/C++ and
Fortran compilers and \maketool{}; PDF rendering uses \latexmk{} and
\pdflatex{}.

\begin{codecell}
\begin{lstlisting}[style=shellcode]
pyamplicol profiling-campaign copy ./performance-campaign \
  --local-amplicol /path/to/AmpliCol \
  --local-madgraph /path/to/MG5_aMC
cd performance-campaign
python steer_performance_campaign.py run \
  --workers 1 --ram-limit 30000000000
python steer_performance_campaign.py refresh-pdf
\end{lstlisting}
\end{codecell}

With no selection filters, this runs the full shipped measurement catalogue,
using one worker and a 30\,GB memory limit.  Repeat the \texttt{run} command
to resume: completed compatible measurements are reused.  The
\texttt{refresh-pdf} command can be called at any time to rebuild
\texttt{pyAmpliCol.pdf} from the available results.  The command
\texttt{run --help} lists the process, multiplicity and execution-mode
selectors, together with concurrency and per-run memory and time limits.

\bibliography{references}

\end{document}